\documentclass[12pt]{article}
\usepackage{graphicx}
\usepackage{latexsym,amsmath,amsfonts,amssymb,color}
\usepackage[latin1]{inputenc}
\usepackage{cite}
\usepackage[american]{babel}
\usepackage{hyperref}
\newcommand{\be}{\begin{equation}}
\newcommand{\ee}{\end{equation}}
\newcommand{\beq}{\begin{equation}}
\newcommand{\eeq}{\end{equation}}
\newcommand{\bea}{\begin{eqnarray}}
\newcommand{\eea}{\end{eqnarray}}

\newcommand{\ba}{\begin{eqnarray}}
\newcommand{\ea}{\end{eqnarray}}
\begin{document}
\baselineskip=15.5pt
\pagestyle{plain}
\setcounter{page}{1}

\def\ie{{\em i.e.},}
\def\eg{{\em e.g.},}
\newcommand{\rc}{\nonumber\\}
\newcommand{\bear}{\begin{eqnarray}}
\newcommand{\eear}{\end{eqnarray}}
\newcommand{\Tr}{\mbox{Tr}}    
\newcommand{\ack}[1]{{\color{red}{\bf Pfft!! #1}}}

\def\a{\alpha}
\def\b{\beta}
\def\c{\gamma}
\def\d{\delta}
\def\eps{\epsilon}           
\def\f{\phi}               
\def\vf{\varphi}  \def\tvf{\hat{\varphi}}
\def\vp{\varphi}
\def\g{\gamma}
\def\h{\eta}
\def\j{\psi}
\def\k{\kappa}                    
\def\l{\lambda}
\def\m{\mu}
\def\n{\nu}
\def\o{\omega}  \def\w{\omega}
\def\p{\pi}
\def\q{\theta}  \def\th{\theta}                  
\def\r{\rho}                                     
\def\s{\sigma}                                   
\def\t{\tau}
\def\u{\upsilon}
\def\x{\xi}
\def\z{\zeta}
\def\pt{\hat{\varphi}}
\def\tt{\hat{\theta}}
\def\lab{\label}
\def\6{\partial}
\def\wg{\wedge}
\def\atanh{{\rm arctanh}}
\def\bpsi{\bar{\psi}}
\def\bt{\bar{\theta}}
\def\bvf{\bar{\varphi}}
\def\W{\Omega}

\numberwithin{equation}{section}

\renewcommand{\theequation}{{\rm\thesection.\arabic{equation}}}


\begin{flushright}
\end{flushright}

\begin{center}

\centerline{\Large {\bf  Anisotropic D3-D5 black holes with unquenched flavors}}

\vspace{8mm}

\renewcommand\thefootnote{\mbox{$\fnsymbol{footnote}$}}
Jos\'e Manuel Pen\'\i n${}^{1,2}$\footnote{jmanpen@gmail.com},\\
Alfonso V. Ramallo${}^{1,2}$\footnote{alfonso@fpaxp1.usc.es}
and Dimitrios Zoakos${}^{3}$\footnote{zoakos@gmail.com}

\vspace{4mm}

${}^1${\small \sl Departamento de  F\'\i sica de Part\'\i  culas} \\
{\small \sl Universidade de Santiago de Compostela} \\
{\small \sl and} \\
${}^2${\small \sl Instituto Galego de F\'\i sica de Altas Enerx\'\i as (IGFAE)} \\
{\small \sl E-15782 Santiago de Compostela, Spain} 
\vskip 0.2cm

${}^3${\small \sl Centro de F\'isica do Porto, Universidade do Porto, \\
Rua do Campo Alegre 687, 4169--007 Porto, Portugal
}

\end{center}

\vspace{8mm}
\numberwithin{equation}{section}
\setcounter{footnote}{0}
\renewcommand\thefootnote{\mbox{\arabic{footnote}}}

\begin{abstract}
We construct a black hole geometry generated by the intersection of $N_c$ color D3- branes and $N_f$ flavor
D5-branes along a 2+1 dimensional subspace.  Working in the Veneziano limit in which $N_f$ is large and distributing   homogeneously  the D5-branes in the internal space, we calculate the solution of the equations of motion of supergravity plus sources which includes the backreaction of the flavor branes.  The solution is analytic and dual to a  2+1 dimensional  defect in a 3+1 dimensional gauge theory,  with $N_f$ massless hypermultiplets living in the defect. The smeared background we obtain can be regarded as the holographic realization of a multilayered system.  We study the thermodynamics of the resulting spatially anisotropic geometry and compute the first and second order transport coefficients for perturbations propagating along the defect.  We find that, in our system,  the dynamics of excitations within a layer can be described by a stack of effective D2-branes. 

\end{abstract}

\newpage
\tableofcontents

\section{Introduction}

The holographic AdS/CFT correspondence \cite{Maldacena:1997re}  has become a useful and powerful tool to study quantum field theories in the strongly coupled regime (see \cite{AdS_CFT_reviews} for reviews). Even if most of the models studied in the holographic framework are very different from the systems found in the phenomenology,  many of the results obtained using them are believed to be universal. To test the universality of these holographic results one should be able to extend the holographic analysis to models  including features present in real life systems.

In this paper we construct a model which allows to explore  the extension of the AdS/CFT correspondence in two directions. First of all, we add dynamical flavors, \ie\ fields transforming in the fundamental representation of the gauge group.  Moreover, our model is dual to a four-dimensional system which is spatially anisotropic since one of the spatial field theory directions of the metric  is distinguished with respect to the other two.  The corresponding geometry is a black hole, \ie\ it has an event horizon, and is based on the D3-D5 brane intersection of type IIB supergravity. The D3-branes are the color branes which, in the absence of D5-branes, generate the $AdS_5\times S^5$ geometry dual to 
$SU(N_c)$  ${\cal N}=4$  super Yang-Mills in $3+1$-dimensions.  The D5-branes are the flavor branes \cite{Karch:2002sh} and are arranged in such a way that they create a $(2+1)$-dimensional,  codimension one,  defect on the worldvolume of the D3-branes.

The field theory dual of this D3-D5 setup is well known.  It was determined some time ago in \cite{DeWolfe:2001pq} (see also \cite{Erdmenger:2002ex,Skenderis:2002vf}).  It consists of a supersymmetric defect theory with (2+1)-dimensional matter hypermultiplets coupled to a (3+1)-dimensional bulk theory.  In the past this D3-D5 setup was extensively studied in the approximation in which the D5-branes are considered as probes in the D3-brane geometry (see, for example,  \cite{Arean:2006pk,Filev:2009xp,Jensen:2010ga,Evans:2010hi,Kristjansen:2012ny,Kristjansen:2013hma,Evans:2014mva,Gomis:2006cu}).  This is the so-called quenched approximation, which corresponds, in the field theory side, to neglecting the quark dynamical effects due to quark loops. This probe brane approach is a good approximation when the number of flavors $N_f$ is much smaller than the number of colors $N_c$.

In this paper we analyze this D3-D5 brane configuration beyond the quenched approximation. To find gravity duals  to unquenched flavor one has to solve the equations of motion of supergravity in the presence of D-brane sources. These sources have Dirac $\delta$-functions and the corresponding Einstein equations are PDE's which are extremely difficult to solve. To overcome this difficulty we follow the proposal of \cite{Bigazzi:2005md} and consider a continuous distribution of D5-brane sources in such a way that there are no $\delta$-functions anymore in our equations of motion.  This approach is accurate only when the number of flavors $N_f$ is large. Actually, it corresponds to the so-called Veneziano limit, in which both $N_c$ and $N_f$ are large and their ratio $N_c/N_f$ is fixed \cite{Veneziano:1976wm}.  This smearing approach has been successfully applied to obtain several geometries dual to flavored systems (see \cite{Nunez:2010sf} for a review and references).  In many cases one gets analytic solutions at the price of modifying the $R$-symmetry of the model (due to the average over different orientations of the flavor branes) and changing the flavor group from $U(N_f)$ to $U(1)^{N_f}$ (the smeared flavor branes are not coincident).

Most of the smeared flavored geometries found in the literature preserve some amount of supersymmetry. Indeed, in these models the preservation of supersymmetry is a crucial guide to find the deformation induced by the flavor branes. However, there are  other solutions which are not supersymmetric and correspond to systems at finite temperature and/or finite baryon density (see \cite{Bigazzi:2009bk,Jokela:2012dw,Bigazzi:2011it,Faedo:2015urf,Faedo:2017aoe}).   For the D3-D5 system we are interested in, the smeared supersymmetric solution has been obtained in \cite{Conde:2016hbg}. In the case of massless quarks the solution is completely analytic and displays a Lifshitz-like anisotropic scaling symmetry. In this paper we find the non-zero temperature generalization of this scaling background. It turns out that adding an event horizon to the geometry of \cite{Conde:2016hbg} is straightforward and amounts to adding a  blackening factor to the metric. This blackening factor has a non-standard power dependence on the radial coordinate due to the spatial anisotropy of the geometry.  

In our background the D5-branes are homogeneously distributed along the internal directions, as well as across the cartesian direction transverse to the defect. Therefore, our gravitational solution should be regarded as the holographic dual of a multilayered system.  The different layers are created by the stack of flavor D5-branes  distributed in parallel two-dimensional planes inside the three-dimensional space. The resulting system has one distinguished direction and thus it is clearly anisotropic. We want to explore its properties for observables living in a single layer and also for those connecting two different layers. We will find that,  non-trivially, the intra-layer dynamics  is the same as that of a stack of  effective D2-branes, which means that strongly coupled 2+1 super Yang-Mills can be used to describe our system.  We will  also be able to study some inter-layer  properties. 

In the condensed matter context it is quite common to have materials with stratified structures containing multiple parallel layers. The possibility of having a holographic top-down model with multiple layers is one of the main motivations for this work.  It is worth recalling  in this respect that  the D3-D5 brane intersection has  been used to model the  quantum Hall effect and  as a holographic model of graphene \cite{Kristjansen:2012ny,Kristjansen:2013hma,Evans:2014mva}.

We will start our analysis by studying  the thermodynamics of the D3-D5 black hole and by computing by different methods the VEV of the stress-energy tensor of the dual theory. This analysis will serve us to characterize the anisotropy of the system  from the holographic perspective. There is an extensive literature on anisotropic holography.  In a by no means exhaustive list, let us mention the articles  \cite{Azeyanagi:2009pr,Mateos:2011tv,Ammon:2012qs,Jain:2014vka,Cheng:2014qia,Banks:2015aca,Roychowdhury:2015cva,Giataganas:2017koz}, where other backgrounds dual to anisotropic theories have been obtained (some of these geometries are also generated by the backreaction of  branes).  We will also be  able to compute the transport coefficients up to second order for perturbations that propagate along the $(2+1)$-dimensional intersection of the D3- and D5-branes.  We will find that these transport coefficients are the same as those of a D2-brane, a result which is  not expected a priori.

It is interesting to recall that localized supergravity solutions for the D3-D5 system have already been found in 
\cite{D'Hoker:2007xy,D'Hoker:2007xz}. These solutions contain cycles with fluxes which can be interpreted as the location of the D5-branes. These D5-branes do not have open string degrees of freedom. This is in contrast to our approach, where the flavor branes are dynamical sources. By smearing these sources we get simpler supersymmetric solutions, which can be easily generalized to construct a black hole.

The organization of the rest of this paper is the following. In section \ref{setup} we present our black hole background, whose thermodynamic  properties are analyzed in section \ref{thermodynamics}. Besides its temperature and entropy, we obtain the chemical potential associated to the D5-brane charge. This allows us to obtain the Helmhotz and Gibbs free energies and find the speed of sound in the directions parallel and orthogonal to the defect. We will check these results by computing the VEV of the stress-energy tensor from the regularized Brown-York tensor of the gravity theory. 

In section \ref{4d_action} we obtain an effective gravitational action for our problem in four-dimensions, which we renormalize holographically by means of a suitable boundary counterterm constructed from a superpotential.  In section \ref{5d_action} we present a five-dimensional gravitational action for our system, which includes a smeared codimension one DBI contribution due to the D5-branes. The regulating boundary term for this action contains a bulk superpotential, as well as a superpotential generated by the flavor branes. We use both the four  and five dimensional regulated actions to calculate the VEV of the stress-energy tensor and to confirm the values obtained in the thermodynamic analysis. In section \ref{hydro} we use the  four-dimensional effective action to compute the transport coefficients in the shear and sound channels.  Finally, in section \ref{conclusions} we summarize our results and discuss possible extensions of our work. The paper is completed with four appendices with details of the calculations presented in the main text.

\section{The D3-D5 black hole}
\label{setup}

In this section we present the brane setup corresponding to our black hole geometry, as well as its metric and forms. More details are provided in appendix \ref{Background_details}. Our background is based on the following array of D3- and D5-branes:
\beq
\begin{array}{ccccccccccl}
 &1&2&3& 4& 5&6 &7&8&9 &  \\
(N_c)\,\,D3: & \times &\times &\times &\_ &\_ & \_&\_ &\_ &\_ &      \\
(N_f)\,\,D5: &\times&\times&\_&\times&\times&\times&\_&\_&\_ &
\end{array}
\label{D3D5intersection}
\eeq
where the $N_c$ D3-branes are color branes and the $N_f$ D5-branes are flavor branes. As it is clear from 
(\ref{D3D5intersection}) the D5-branes create a (2+1)-dimensional defect in the (3+1)-dimensional bulk gauge theory. In general, the directions 4-9 correspond to a Sasaki-Einstein cone, with the D3-branes located at the tip of the cone. For concreteness we will consider here the case in which the D3-branes are in flat space and,  therefore, the base of the cone will be just the five-sphere ${\mathbb S}^5$.

The ten-dimensional metric of our geometry in Einstein frame has the factorized form:
\beq
ds^2_{10}\,=\,ds^2_5\,+\,d\hat s_5^2\,\,,
\label{10d_flavored_metric}
\eeq
where $ds^2_5$ is:
\beq
ds^2_5\,=\,{r^2\over R^2}\,\Big[-b\,(dx^0)^2\,+\,(dx^1)^2\,+\,(dx^2)^2\,+\,e^{-2\phi}\,
\,(dx^3)^2\,\Big]\,+\,
R^2\,{dr^2\over b\, r^2}\,\,,
\label{black-anisotropic}
\eeq
where  $R$ is a constant radius and $b=b(r)$ is the blackening factor, given by:
\beq
b\,=\,1\,-\,\Big({r_h\over r}\Big)^{{10\over 3}}\,\,,
\label{blackening_factor_r}
\eeq
with $r_h$ being the horizon radius. The function $\phi$ multiplying the metric  (\ref{black-anisotropic})  along the $x^3$ direction is the type IIB supergravity dilaton, which is not constant due to the presence of the D5-branes. The running of $\phi$ characterizes the anisotropy introduced by the flavor branes in the $(3+1)$-dimensional gauge theory.

The metric $d\hat s_5^2$ in (\ref{10d_flavored_metric}) corresponds to the internal part of the 10d geometry. As  in the smeared solution of \cite{Conde:2016hbg} this internal metric is just a deformed  ${\mathbb S}^5$. This deformation can be easily described when the ${\mathbb S}^5$ is represented as 
a $U(1)$ bundle over ${{\mathbb C}{\mathbb P}^2}$: the deformation is just a squashing of the $U(1)$ fiber relative to the ${{\mathbb C}{\mathbb P}^2}$ base. Actually, the internal part of our metric is:
\beq
d\hat s_5^2\,=\,\bar R^2\,\big[
\,ds^2_{{\mathbb C}{\mathbb P}^2}+{9\over 8}\,(d\tau+A)^2 \big]\,\,,
\label{compact_flavored_metric}
\eeq
where $\bar R^2$ is a constant related to the radius $R$ as: 
\beq
\bar R^2\,=\,{9\over 8}\,R^2\,\,.
\eeq

Our backreacted background is a solution of the equations of motion derived from the total action of the system, which is the sum of the type IIB supergravity action and of the action of the D5-branes:
\beq
S\,=\,S_{IIB}\,+\,S_{branes}\,\,.
\label{total_10d_action}
\eeq
The action of type IIB supergravity in Einstein frame is:
\beq
S_{IIB} = \frac{1}{2\kappa_{10}^2} \left[ \int d^{10}x\, \sqrt{-g} \left( R - \frac{1}{2} \partial_{\mu} \phi \partial^{\mu} \phi \right) -  \int \left( \frac{1}{2}\,e^{\phi} F_{3} \wedge *F_{3} \,+\,{1\over 4}\,
F_{5} \wedge *F_{5} \right) \right]\,\,,
\label{typeIIB_action}
\eeq
while the action of the branes is given by the  sum of DBI and WZ terms:
\beq
S_{branes}\,=\,-T_5\,\sum_{N_f}\,\Bigg(\,
\int_{{\cal M}_6}\,d^6\xi\,e^{{\phi\over 2}}\,\sqrt{-\hat g_6}\,-\,\int_{{\cal M}_6}\ \hat C_6\Bigg)\,\,,
\label{S_branes}
\eeq
where $T_5$ is the tension of the D5-brane ($1/T_5\,=\,(2\pi)^5\,g_s\,(\,\alpha'\,)^3$), $\hat g_6$ is the determinant of the induced metric on the worldvolume  ${\cal M}_6$ and $\hat C_{(6)}$  is the  pullback to ${\cal M}_6$ of the  RR six-form potential of the type IlB theory.  In (\ref{typeIIB_action})  we have only included the RR three- and five-forms $F_3$ and $F_5$, which are the only non-trivial ones for our D3-D5 geometry.

The stack of color D3-branes induces a self-dual RR five-form $F_5$ of the type:
\beq
F_5\,=\,K(r)\,\big(1+*\big)\,d^4x\wedge dr\,\,,
\label{F5_ansatz}
\eeq
where $K=K(r)$ is a function of the radial variable whose explicit expression can be found in appendix \ref{Background_details} (eq. (\ref{F5_sol})).  Moreover, the $N_f$ flavor D5-branes act as a source of the RR three-form $F_3$ through the WZ term of the action (\ref{S_branes}).

 In the smearing approach, valid when $N_f$ is large, we substitute the discrete distribution of flavor branes by a continuous distribution with the appropriate normalization, in such a way that the smearing amounts to performing the substitution:
\beq
\sum^{N_f}\,\int_{{\cal M}_6}\,\hat C_{(6)}\,\,
\Longrightarrow\,\,
\int_{{\cal M}_{10}}\,\Xi\wedge C_{(6)}\,\,,
\eeq
where  $\Xi$  is a four-form (the so-called smearing form),  with components along the  directions orthogonal to the worldvolume of the flavor branes, which characterizes the charge distribution of the flavor branes.   As shown in \cite{Conde:2016hbg} this WZ coupling induces the following violation of Bianchi identity of  $F_{(3)}$:
\beq
dF_3\,=\,2\,\kappa_{10}^2\,T_5\,\Xi\,\,.
\eeq
The detailed form of $F_3$ and $\Xi$ in terms of differentials of the coordinates is given in appendix \ref{Background_details}  (see (\ref{F3_ansatz}) and (\ref{dF_3})).  It is important to notice  that $\Xi$ does not depend on $x^3$ (it only depends on $dx^3$), which means that we are homogeneously distributing  our flavor branes in the $x^3$ direction and, therefore, we can regard our setup as dual to a multilayer system. Moreover,  $\Xi$  is also independent of the radial coordinate $r$, as expected for a charge distribution corresponding to massless quarks.  The radii $R$ and $\bar R$ depend on the number of color branes $N_c$. Indeed, they can be written as:
\beq
R^4\,=\,{256\over 1215}\,Q_c\,\,,
\qquad\qquad
\bar R^4\,=\,{4\over 15}\,Q_c\,\,,
\label{R_Qc}
\eeq
where $Q_c$ is proportional to $N_c$ and given by:
\beq
Q_c\,=\,{(2\pi)^4\,g_s\,\alpha'{}^{\,2}\,N_c\over {\rm Vol}({\cal M}_5)}\,=\,16\,\pi\,g_s\,\alpha'{}^{\,2}\,\,N_c\,\,.
\label{Qc}
\eeq
In what follows we will take $g_s=\alpha'=1$. Moreover, $F_3$ and the dilaton $\phi$ depend on the quantity $Q_f\sim N_f$, as shown in (\ref{F3_ansatz}) and (\ref{dilaton_sol}). The precise relation between $Q_f$ and $N_f$ is written in (\ref{Qf_Nf}). It is important to point out that our solution is not analytic in $N_f$, which means that we cannot take the unflavored  limit $N_f=0$ and recover the isotropic $AdS_5\times {\mathbb S}^5$ background.\footnote{One can take this $N_f=0$ limit in the equations of motion but not in their particular solution corresponding to our background.}

When $r_h=0$ (and $b=1$) our  solution is supersymmetric, as shown in \cite{Conde:2016hbg}, and can be found by solving a set of first-order BPS equations. This supersymmetric solution is invariant under a set of Lifshitz-like anisotropic scale transformations in which the $x^3$ coordinate transforms with an anomalous exponent $z=3$ (see \cite{Conde:2016hbg}  for further details about this scaling symmetry).

In order to explore the physical consequences of the anisotropy of our background, we have computed in appendix \ref{WL-EE} the potential energy, at zero temperature, for a quark-antiquark pair, following the holographic prescription of refs. \cite{Maldacena:1998im,Rey:1998ik}. We have considered the cases in which the charges are in the same layer (\ie\ when they have the same value of $x^3$) and when they are separated along $x^3$. Let us summarize here the results. The intra-layer potential takes the form:
\beq
V_{q\bar q}\sim {N_c^{{2\over 3}}\over N_f^{{1\over 3}}}\,{1\over d_{\parallel}^{{4\over 3}}}\,\,,
\label{V_parallel}
\eeq
where $d_{\parallel}$ is the $q\bar q$ distance in the $x^1 x^2$ plane. Moreover, for charges with the same values of $(x^1, x^2)$  and separated a distance $d_{\perp}$ along the coordinate $x^3$, we obtain:
\beq
V_{q\bar q}\sim {N_c^2\over N_f^3}\,{1\over d_{\perp}^4}\,\,.
\label{V_perp}
\eeq
The different behaviors (\ref{V_parallel}) and (\ref{V_perp}) gives us a measure of the effects of the anisotropy on physical observables. Another effect of this anisotropy is encoded in the entanglement entropies for slab regions and their complements at zero temperature. For a slab with a finite width $l_{\parallel}$ in the plane, the entanglement entropy behaves as (see appendix \ref{WL-EE}  for details):
\beq
S_{\parallel}\sim {N_f^{{2\over 3}}\,N_c^{{5\over 3}}\over l_{\parallel}^{{4\over 3}}}\,\,,
\label{S_parallel}
\eeq
whereas if the slab has a finite width $l_{\perp}$ along $x^3$ we get:
\beq
S_{\perp}\sim {N_c^4\over N_f^4}\,{1\over l_{\perp}^6}\,\,.
\label{S_perp}
\eeq
Eqs. (\ref{S_parallel}) and (\ref{S_perp})  contain information about the quantum correlations of the model. In particular, the dependence of the entropies on the length determines the critical behavior of the mutual information. Interestingly, $S_{\parallel}$ depends on $N_c$ and $l_{\parallel}$ as in the case of a D2-brane. We will find several times in this paper this equivalence of the intra-layer physics with the one corresponding to an effective D2-brane.

When $r_h\not=0$ our solution has a horizon and becomes a black hole with a non-zero temperature. In this case one can show that it solves the Einstein equations with sources that follow from the action (\ref{total_10d_action}). In particular the DBI term of (\ref{S_branes}) contributes to the energy-momentun tensor and, as already mentioned, the WZ term induces a violation of the Bianchi identity of $F_3$.  In the next section we explore the thermodynamic properties of this black hole.

\section{Thermodynamics of the black hole}
\label{thermodynamics}

Let us now work out the thermodynamics of the black hole presented in the previous section. First of all, we recall that  the temperature $T$ is given by the general formula:
\beq
T\,=\,{1\over 2\pi}\,\,
\Big[\,{1\over \sqrt{g_{rr}}}\,\,{d\over dr}\,
\Big(\,\sqrt{\,-g_{x^0 x^0}}\,\Big)\,
\Big]_{r=r_h}\,\,,
\eeq
which leads to the following relation between $T$ and the horizon radius $r_h$:
\beq
T\,=\,{5\,r_h\over 6\pi\, R^2}\,\,.
\label{T_rh_R}
\eeq
Using (\ref{R_Qc}) we can recast this relation in terms of $Q_c$ as:
\beq
r_h\,=\,{2^5\,\pi\over 3^{{3\over 2}}\,5^{{3\over 2}}}\,
Q_c^{{1\over 2}}\,T\,\,.
\label{rh_T_Q_c}
\eeq
The entropy density $s$ is given by the Bekenstein-Hawking formula:
\beq
s\,=\,{2\pi\over \kappa_{10}^2}\,\,{A_8\over V_3}\,\,,
\eeq
where $A_8$ is the volume at the horizon of the eight-dimensional space orthogonal to $t$ and $r$ and $V_3$ is the infinite constant volume of the three-dimensional Minkowski directions. For our black hole geometry we get:
\beq
{A_8\over V_3}\,=\,2^{-{11\over 3}}\,3^{{17\over 6}}\,5^{-{1\over 2}}\,\pi^3\,
Q_f^{{2\over 3}}\,Q_c^{{1\over 2}}\,r_h^{{7\over 3}}\,\,.
\eeq
After using (\ref{rh_T_Q_c}) to relate $r_h$ and $T$, we arrive at:
\beq
s\,=\,{2^3\over 5^4\,3^{{2\over 3}}\,\pi^{{2\over 3}}}\,
Q_f^{{2\over 3}}\,Q_c^{{5\over 3}}\,T^{{7\over 3}}\,\,.
\label{s_Qf_Qc}
\eeq
Notice the fractional powers of $Q_c$ and $Q_f$ in (\ref{s_Qf_Qc}), which mean that $s$ has a non-standard  dependence on $N_c$ and $N_f$. To explore further this dependence,  let us  rewrite (\ref{s_Qf_Qc})  in terms of $N_c$ and $N_f$. With this purpose we use the relations
(\ref{Qc}) and (\ref{Qf_Nf}),  from which we get that the combination appearing in (\ref{s_Qf_Qc}) is given by:
\beq
Q_f^{{2\over 3}}\,Q_c^{{5\over 3}}\,=\,
{256\over 3\,3^{{2\over 3}}}\,\pi^{{7\over 3}}\,
N_f^{{2\over 3}}\,N_c^{{5\over 3}}\,\,,
\label{Qf_Qc_powers}
\eeq
and the entropy density can be written as:
\beq
s\,=\,\alpha_s\,N_f^{{2\over 3}}\,N_c^{{5\over 3}}\,
T^{{7\over 3}}\,\,,
\label{entropy_density}
\eeq
where $\alpha_s$ is the following numerical coefficient:
\beq
\alpha_s\,=\,{2048\over 5625}\,{\pi^{{5\over 3}}\over 3^{{1\over 3}}}\,\approx\,1.701\,\,.
\label{alpha_s}
\eeq

The ADM energy of the background is given by the standard equation:
\beq
E_{ADM}\,=\,-{1\over \kappa_{10}^2}\,
\sqrt{|g_{tt}|}\,\,\int_{{\cal M}_{t,r_{\infty}}}\sqrt{\det g_8}\,\,
(\,K_T\,-\,K_0\,)\,\,,
\label{ADM_energy}
\eeq
where the symbols $K_T$ and $K_0$ denote the extrinsic curvatures of the eight-dimensional subspace within the nine-dimensional (constant time) space, at finite and zero temperature, respectively. For an arbitrary hypersurface the extrinsic curvature $K$ is given by:
\beq
K\,=\,{1\over \sqrt{\det g_9}}\,\partial_{\mu}\,
\Big(\,\sqrt{\det g_9}\,\,n^{\mu}\,\Big)\,\,,
\eeq
with $n^{\mu}$ being a normalized vector perpendicular to the surface. For a constant $r$ hypersurface, we have:
\beq
n^{\mu}\,=\,{1\over \sqrt{g_{rr}}}\,\delta_{r}^{\mu}\,\,.
\eeq
For our geometry it is straightforward to prove that:
\beq
K\,=\,{7\over 3R}\,\sqrt{b}\,\,,
\eeq
where $b$ is the blackening factor (\ref{blackening_factor_r}).  From this result it follows that:
\beq
K_T\,=\,{7\over 3R}\,\sqrt{1\,-\,\Big({r_h\over r}\Big)^{{10\over 3}}}\,\,,
\qquad\qquad
K_0\,=\,{7\over 3R}\,\,,
\eeq
and thus the difference of the extrinsic curvatures appearing in (\ref{ADM_energy}) is:
\beq
K_T\,-\,K_0\,\approx\,-{7\over 6R}\,\Big({r_h\over r}\Big)^{{10\over 3}}\,\,,
\qquad\qquad
(r\to\infty)\,\,.
\eeq
The energy density $\epsilon$ can now be easily computed, with the result:
\beq
\epsilon\,=\,{E_{ADM}\over V_3}
\,=\,{7\over 10}\,\alpha_s\,
N_f^{{2\over 3}}\,N_c^{{5\over 3}}\,
T^{{10\over 3}}\,=\,
\beta_s\,Q_f^{{2\over 3}}\,Q_c^{{5\over 3}}\,T^{{10\over 3}}\,\,,
\label{energy_density}
\eeq
where $\alpha_s$ is the numerical coefficient (\ref{alpha_s}) and we have introduced a new numerical factor $\beta_s$, given by:
\beq
\beta_s\,=\,{28\over 3125\,(3\pi)^{{2\over 3}}}\,\,.
\label{beta_s}
\eeq
Notice that the entropy density (\ref{s_Qf_Qc}) can be rewritten as:
\beq
s\,=\,{10\over 7}\,\beta_s\,Q_f^{{2\over 3}}\,Q_c^{{5\over 3}}\,T^{{7\over 3}}\,\,.
\label{sQs_beta}
\eeq
The free energy density $f$ in the canonical ensemble  is defined as:
\beq
f\,=\,\epsilon\,-\,T\,s\,\,.
\label{f_definition}
\eeq
By using (\ref{entropy_density}) and (\ref{energy_density}) we readily obtain:
\beq
f\,=\,-{3\over 10}\,\alpha_s\,
N_f^{{2\over 3}}\,N_c^{{5\over 3}}\,
T^{{10\over 3}}\,=\,-{3\over 7}\,\beta_s\,Q_f^{{2\over 3}}\,Q_c^{{5\over 3}}\,T^{{10\over 3}}\,\,.
\label{free_energy_density}
\eeq

To explore the complete thermodynamics of the system it is convenient to consider the situation in which the number of flavor D5-branes can change. In our setup this number of flavor branes is determined by $Q_f$. Therefore, we allow $Q_f$ to vary and we will introduce the chemical potential $\Phi$, conjugate to $Q_f$.  The first law of thermodynamics for these variables becomes:
\beq
d\epsilon\,=\,T\,ds\,+\,\Phi\, dQ_f\,\,.
\eeq
Clearly, the chemical potential $\Phi$ measures the energy cost of introducing additional flavor branes in the system.  After performing the Legendre transform as in (\ref{f_definition}), we can write the variation of the  free energy $f$ in the canonical ensemble as:
\beq
df\,=\,-s\,dT\,+\,\Phi\,dQ_f\,\,.
\label{df_dT_dQ}
\eeq
It follows immediately  from (\ref{df_dT_dQ}) that $s$ and $\Phi$ are given by the following partial derivatives of $f$:
\beq
s\,=\,-\Bigg({\partial f\over \partial T}\Bigg)_{Q_f}\,\,,
\qquad\qquad
\Phi\,=\,\Bigg({\partial f\over \partial Q_f}\Bigg)_{T}\,\,.
\label{partial_der_f}
\eeq
By using (\ref{free_energy_density}),  it is now straightforward to compute the partial derivative of $f$ with respect to $T$ and check the first equation in (\ref{partial_der_f}).  Moreover, by computing the derivative of (\ref{free_energy_density}) with respect to $Q_f$ we obtain the expression of the chemical potential $\Phi$:
\beq
\Phi\,=\,-{2\over 7}\,\beta_s\,Q_f^{-{1\over 3}}\,Q_c^{{5\over 3}}\,T^{{10\over 3}}\,\,.
\label{Phi_Qs_beta}
\eeq
The Gibbs free energy, \ie\ the thermodynamic potential in the grand canonical ensemble, is defined as:
\beq
g\,=\,f\,-\,\Phi\,Q_f\,\,.
\label{g_def}
\eeq
Plugging (\ref{free_energy_density}) and (\ref{Phi_Qs_beta}) on the right-hand side of (\ref{g_def}) we get the value of $g$ for our system:
\beq
g\,=\,-{1\over 7}\,\beta_s\,Q_f^{{2\over 3}}\,Q_c^{{5\over 3}}\,T^{{10\over 3}}\,\,.
\eeq
As argued in \cite{Mateos:2011tv} (see also \cite{Caldarelli:2010xz}), the two thermodynamic potentials $f$ and $g$ are related to the pressure in the $x^1x^2$ plane ($p_{xy}$)  and in the $x^3$ direction ($p_z$)  as:
\beq
f\,=\,-p_{xy}\,\,,
\qquad\qquad
g\,=\,-p_z\,\,.
\eeq
To demonstrate these identifications of the free energies with the pressures one should take into account the extensivity of the energy and the anisotropic character of our system (details can be found in \cite{Mateos:2011tv}).  In our system these pressures are thus  given by:
\beq
p_{xy}\,=\,{3\over 7}\,\beta_s\,Q_f^{{2\over 3}}\,Q_c^{{5\over 3}}\,T^{{10\over 3}}\,=\,{3\over 7}\,\epsilon\,\,,
\qquad\qquad
p_{z}\,=\,{1\over 7}\,\beta_s\,Q_f^{{2\over 3}}\,Q_c^{{5\over 3}}\,T^{{10\over 3}}\,=\,{1\over 7}\,\epsilon\,\,.
\label{ps_Qs_epsilon}
\eeq
The speeds of sound  along the $x^1x^2$   and $x^3$ are defined as:
\beq
v_{xy}^2\,=\,\Big({\partial p_{xy}\over \partial \epsilon}\Big)_{Q_f}\,\,,
\qquad\qquad
v_{z}^2\,=\,\Big({\partial p_{z}\over \partial \epsilon}\Big)_{Q_f}\,\,. 
\label{vs_def}
\eeq
Using (\ref{ps_Qs_epsilon}) we can readily evaluate the derivatives on the right-hand side of (\ref{vs_def}), with the result:
\beq
v_{xy}^2\,=\,{3\over 7}\,\,,
\qquad\qquad
v_{z}^2\,=\,{1\over 7}\,\,,
\label{speeds_of_sound}
\eeq
to be compared with the value $v_s^2=1/2$ for a 2d CFT and $v_s^2=1/3$ for a 3d CFT.\footnote{The 
speed of sound for a Dp-brane is $v_s^2={5-p\over 9-p}$. Therefore $v_{xy}$ coincides with the speed of sound of a D2-brane.}

The pressure difference is a manifestation of the anisotropy of the system and is measured by the non-vanishing chemical potential. Actually, it is straightforward to verify that, for our system, one has:
\beq
p_z\,-\,p_{xy}\,=\,\Phi\,Q_f\,\,.
\label{delta_p}
\eeq
Moreover, we have the following equation of state:
\beq
\epsilon\,=\,2\,p_{xy}\,+\,p_z\,\,.
\label{e_o_s}
\eeq
By combining (\ref{delta_p}) and (\ref{e_o_s}) we can obtain the two pressures as  functions of $\epsilon$ and $Q_f$:
\beq
p_{xy}\,=\,{1\over 3}\,\epsilon\,-\,{1\over 3}\,\Phi\,Q_f\,\,,
\qquad\qquad
p_z\,=\,{1\over 3}\,\epsilon\,+\,{2\over 3}\,\Phi\,Q_f\,\,.
\eeq
It is also easy  to relate the different quantities to  the entropy:
\beq
\epsilon={7\over 10}\,Ts\,\,,\qquad
f\,=\,-{3\over 10}\,Ts\,\,,\qquad
g\,=\,-{1\over 10}\,Ts\,\,,\qquad
\Phi\,Q_f\,=-{1\over 5}\,Ts\,\,.
\eeq
From these equations one can show that the following relation holds:
\beq
\epsilon\,=\,{3\over 4}\,Ts\,+\,{1\over 4}\,\Phi\,Q_f\,\,,
\eeq
as well as the so-called  Gibbs-Duhem relations:
\beq
\epsilon\,+\,p_{xy}\,=\,Ts\,\,,
\qquad\qquad
\epsilon\,+\,p_{z}\,=\,Ts\,+\,\Phi\,Q_f\,\,.
\eeq
Finally, the heat capacity is:
\beq
c_v\,=\,\partial_T\,\epsilon\,=\,{7\over 3}\,\alpha_s\,N_f^{{2\over 3}}\,N_c^{{5\over 3}}\,
T^{{7\over 3}}\,=\,{10\over 3}\,\beta_s\,Q_f^{{2\over 3}}\,Q_c^{{5\over 3}}\,
T^{{7\over 3}}\,\,.
\eeq
To get some insight on the nature of our solution, let us analyze the dependence of the entropy density $s$ on $N_c$, $N_f$ and $T$  and let us compare it with some known results for other gravity duals. It follows from (\ref{sQs_beta}) that $s$ behaves with the temperature as $s\sim T^{{7\over 3}}$. For a Dp-brane background $s\sim T^{{9-p\over 5-p}}$ \cite{Itzhaki:1998dd}. Taking $p=2$ in this last formula we obtain the same behavior as in (\ref{sQs_beta}). This is an indication that our geometry is related to the one generated by D2-branes. Actually, if we define $\lambda$ as:
\beq
\lambda\,=\,{N_c\over N_f^2}\,\,,
\label{coupling_scaling}
\eeq
then the  entropy density (\ref{sQs_beta}) can be written as:
\beq
s\sim N_c^2\,\lambda^{-{1\over 3}}\,T^{{7\over 3}}\,\,,
\label{s_D2}
\eeq
which is exactly the form of the entropy of a D2-brane black hole if $\lambda$ is interpreted as a 't Hooft coupling\cite{Itzhaki:1998dd}.\footnote{Equivalently, if we define the temperature-dependent effective dimensionless coupling as $\lambda_{eff}(T)=\lambda/T$, the entropy  density (\ref{s_D2}) can be written as
$s\sim N_c^2\,[\lambda_{eff}(T)]^{-{1\over 3}}\,T^2$.  We are grateful to Javier Tarr\'\i o for suggesting this interpretation of our entropy formula.}  In the case of a stack of $N_c$ D2-branes, realizing $2+1$ dimensional super Yang-Mills, the
't Hooft coupling is $\lambda=g^2_{YM}\,N_c$ ($\lambda=N_c$ in our units). Our result suggests that, in our flavored system, the relevant scaling of the coupling with $N_c$ and $N_f$ is the one written in (\ref{coupling_scaling}). 
Notice that having a ratio of the numbers of color and flavors as parameter is very natural in a  limit of the Veneziano type. Notice also \cite{Jensen:2013lxa} that the dimensionless parameter controlling the backreaction of the flavor D5-branes is $\kappa_{10}^2\,N_f\,T_{D5}\,R^{-2}\sim N_f/\sqrt{N_c}$. In this parameter $N_c$ and $N_f$ scale precisely as in (\ref{coupling_scaling}). 

 The matching  we found of the entropy with the one corresponding to a D2-brane is an indication that the dynamics in the $x^1 x^2$ plane (at least its deviation from conformality) is governed by a $2+1$ dimensional super Yang-Mills theory in the strongly coupled regime. The value of the speed of sound  $v_{xy}$ found above points in the same direction. In section \ref{hydro} we will confirm this fact by computing the hydrodynamic transport coefficients for perturbations propagating in the  $x^1 x^2$ plane.  Actually, there is a direct way to relate our setup to a system of D2-branes. Indeed, by performing a T-duality transformation along the $x^3$ direction we can convert our D3-D5 solution into a D2-D6 geometry, in which the D2's are the color branes and the D6's are the flavor branes. In this D2-D6 solution the $x^3$ direction is now a  distinguished coordinate transverse to the color branes.  The corresponding ten-dimensional metric of type IIA supergravity in the Einstein frame takes the form:
 \bear
&&ds^2_{IIA}\,=\, \Big({4\,Q_f\over 3}\Big)^{{1\over 4}}\,{r^{{9\over 4}}\over R^{{5\over 2}}}
\Bigg[-b\,(dx^0)^2\,+\,(dx^1)^2\,+\,(dx^2)^2\,+\,{R^4\over r^4}{dr^2\over b}\,+\,\rc\rc
&&\qquad\qquad\qquad \qquad
+\,{9\over 8}{ R^4\over r^2}\,\Big({(d\bar x^3)^2\over r^{{4\over 3}}}\,+\,
ds^2_{{\mathbb C}{\mathbb P}^2}+{9\over 8}\,(d\tau+A)^2 \Big)\Bigg]\,\,,
\eear
 where $b=b(r)$ is the blackening factor (\ref{blackening_factor_r}) and the coordinate $\bar x^3$ is related to the original cartesian coordinate $x^3$ by the following rescaling:
  \beq
 \bar x^3\,=\,\Big({4\sqrt{2}\over 9\,Q_f}\Big)^{{1\over 3}}\,x^3\,\,.
 \eeq
 Notice that the D2-branes in this D2-D6 solution are smeared in $x^3$, since none of the functions of the metric depends on this coordinate. 
 
 This type IIA background is  also endowed with a running dilaton $\phi_{IIA}$,  as well as RR two- and four-forms,  given by:
 \bear
&& e^{2\phi_{IIA}}\,=\,\Big({3\over 4\,Q_f}\Big)^{{7\over 3}}\,R^2\,r^{{1\over 3}}\,\,,\rc\rc
&&F_2\,=\,Q_f\,{\rm Im}\,(\hat\Omega_2)\,\,,\rc\rc
&&F_4\,=\,{20\over 3}\Big({2\,Q_f^2\over 9}\Big)^{{1\over 3}}\,{r^{{7\over 3}}\over R^4}\,
dr\wedge dx^0\wedge dx^1\wedge dx^2\,\,.
\eear

\subsection{Stress-energy tensor}
\label{thermo_T_mu_nu}

The energy density and the pressures of our model can also be obtained by calculating  the holographic stress-energy tensor. We will compute this tensor by using several methods and we will check that one gets the same results as those we obtained in the previous subsection by using anisotropic thermodynamics. In this subsection we will compute the VEV of the stress-energy tensor from the Brown-York tensor at the boundary, following the prescription of \cite{Balasubramanian:1999re}. In sections \ref{4d_action} and \ref{5d_action} we will dimensionally reduce our ten-dimensional theory and will calculate the stress-energy tensor by holographic renormalization, after adding suitable boundary terms to the reduced actions.

The Brown-York tensor of the ten-dimensional gravity theory is:
\beq
\tau_{ij}\,=\,{1\over \kappa_{10}^2}\,\big(K_{ij}\,-\,K\,\gamma_{ij}\big)\,\,,
\label{BY_tensor_def}
\eeq
where $\gamma_{ij}$ is the induced metric at a $r={\rm constant}$ surface, $K_{ij}$ is the extrinsic curvature of the surface and $K=\gamma^{ij}\,K_{ij}$.  The VEV of the stress-energy tensor of  the dual theory is related to  the Minkowski components of the  Brown-York tensor at the boundary \cite{Balasubramanian:1999re}:
\beq
\langle T^{\mu}_{\,\,\,\,\nu}\rangle\,=\,V_{SE}\,\sqrt{-\gamma_{Min}}\,\,\tau^{\mu}_{\,\,\,\,\nu\,}\Big|_{\,reg\,,\,r_{\Lambda}\to\infty}\,\,,
\label{em_Tensor-BY}
\eeq
where $V_{SE}$ is the volume for the compact 5d part of the metric, which  for the $S^5$ is 
\beq
V_{SE}\,=\,\Big({9\,\pi\over 8}\Big)^3\,R^5\,\,.
\eeq
In (\ref{em_Tensor-BY}) $\gamma_{Min}$  is the determinant of the Minkowski part of the induced metric. The right-hand side of (\ref{em_Tensor-BY}) is divergent at the UV boundary. We will give below a precise prescription to eliminate this divergence. 

The extrinsic curvature tensor $K_{ij}$ can be obtained from the covariant expression:
\beq
K_{ij}\,=\,-{1\over 2}\,\big(\nabla_{i}\,n_j\,+\,\nabla_{j}\,n_i\big)\,\,,
\eeq
where $n_i$ are the components of the normal vector to the $r={\rm constant}$ surface  ( $n^i\,n_i\,=\,1$). In a diagonal metric as the one we have in (\ref{black-anisotropic}), the vector $n_i$ is given by:
\beq
n_i\,=\,\sqrt{g_{rr}}\,\,\delta_i^r\,\,.
\eeq
Let us now introduce the notation:
\bear
&&g_{x^0 x^0}\,\equiv\,-k_1^2\,=\,-{r^2\over R^2}\,b\,\,,
\qquad\qquad\qquad\,\,\,\,\,
g_{x^1 x^1}\,=\,g_{x^2 x^2}\,\equiv\,k_2^2\,=\,{r^2\over R^2}\,\,,\rc\rc
&&g_{x^3 x^3}\,\equiv\,k_3^2\,=\,{1\over R^2}\,\Big({4 Q_f\over 3}\Big)^{{4\over 3}}\,\,r^{{2\over 3}}\,\,,
\qquad\qquad
g_{rr}\,\equiv\,k_r^2\,=\,{R^2\over r^2\,b}\,\,,
\eear
where we are assuming that the metric is given by (\ref{black-anisotropic}) and (\ref{compact_flavored_metric}).  With these notations, we have:
\beq
\sqrt{-\gamma_{Min}}\,=\,k_1\,k_2^2\,k_3\,=\,{1\over R^4}\,\Big({4 Q_f\over 3}\Big)^{{2\over 3}}\,
r^{{10\over 3}}\,b^{{1\over 2}}\,\,,
\label{det_Min}
\eeq
and it is straightforward to compute the components of the extrinsic curvature along the Minkowski directions. The non-vanishing components are:
\bear
&&K_{x^0 x^0}\,=\,{k_1\,k_1'\over k_r}\,\,,
\qquad\qquad
K_{x^1 x^1}\,=\,K_{x^2 x^2}=-{k_2\,k_2'\over k_r}\,\,,\rc\rc
&&
K_{x^3 x^3}\,=\,-{k_3\,k_3'\over k_r}\,\,,
\qquad\qquad
K\,=\,-{1\over k_r}\,\partial_r\log\big(k_1\,k_2^2\,k_3\big)\,\,.
\eear
Plugging these  results in (\ref{BY_tensor_def}) we get explicitly the non-zero  components of the Brown-York tensor:
\bear
&&\tau^{x^0}_{\,\,\,\,x^0}\,=\,{1\over \kappa_{10}^2}\,
{1\over k_r}\,\partial_r\log \big(k_2^2\,k_3\big)\,=\,{
1\over \kappa_{10}^2\,R}\,{7\over 3}\,b^{{1\over 2}}\,\,,\rc\rc
&&\tau^{x^1}_{\,\,\,\,x^1}\,=\,\tau^{x^2}_{\,\,\,\,x^2}\,=\,{1\over 2\,\kappa_{10}^2\,R}\,
r\,b^{{1\over 2}}\,\partial_r\log\big(r^{{14\over 3}}\,b\big)\,=\,
{1\over \kappa_{10}^2\,R}\,{1\over 3\,b^{{1\over 2}}}\,\,
\Big[7-2 \Big({r_h\over r}\Big)^{{10\over 3}}\Big]\,\,,\rc\rc
&&\tau^{x^3}_{\,\,\,\,x^3}\,=\,{1\over 2\,\kappa_{10}^2\,R}\,
r\,b^{{1\over 2}}\,\partial_r\log\big(r^{6}\,b\big)\,=\,
{1\over \,\kappa_{10}^2\,R}\,{1\over  b^{{1\over 2}}}\,
\Big[3\,-\,{4\over 3}\,\Big({r_h\over r}\Big)^{{10\over 3}}\Big]\,\,.
\label{BY_tensor_exp}
\eear

Let us now specify the regulating procedure we will employ to compute $\langle T^{\mu}_{\,\,\,\,\nu}\rangle$.  Since we are interested in matching the thermodynamic values found above, it is enough to subtract the zero temperature supersymmetric value, as it  was done in \cite{Bigazzi:2009bk} for the D3-D7 system.   More concretely, we will take $\langle T^{\mu}_{\,\,\,\,\nu}\rangle$ to  be given by:
\beq
\langle T^{\mu}_{\,\,\,\,\nu}\rangle\,=\,V_{SE}\lim_{r_{\Lambda}\to\infty}\,
\Bigg[\sqrt{-\gamma_{Min}}\,\,\tau^{\mu}_{\,\,\,\,\nu}\,-\,b^{{1\over 2}}\,
\lim_{r_h\to 0}\,\big(\sqrt{-\gamma_{Min}}\,\,\tau^{\mu}_{\,\,\,\,\nu}\big)\Bigg]_{r=r_{\Lambda}}\,\,,
\eeq
where the $b^{{1\over 2}}$ factor is introduced to match the geometries at the cutoff. Using (\ref{det_Min}) and (\ref{BY_tensor_exp}) we get that the only non-zero components of $\langle T^{\mu}_{\,\,\,\,\nu}\rangle$ are:
\beq
\langle T^{x^0}_{\,\,\,\,x^0} \rangle\,=\,-\epsilon\,\,,
\qquad\qquad
\langle T^{x^1}_{\,\,\,\,x^1} \rangle\,=\,\langle T^{x^2}_{\,\,\,\,x^2} \rangle\,=\,{3\,\epsilon\over 7}\,\,,
\qquad\qquad
\langle T^{x^3}_{\,\,\,\,x^3} \rangle\,=\,{\epsilon\over 7}\,\,,
\eeq
where $\epsilon$ is the ADM energy density (\ref{energy_density}).  Equivalently, we can write the VEV of the stress-energy tensor as:
\beq
\langle T^{\mu}_{\,\,\,\,\nu}\rangle\,=\,{\rm diag}\big(-\epsilon\,,\, p_{xy}\,,\, p_{xy}\,, \,p_z\big)\,\,,
\label{T_values_10d}
\eeq
where $p_{xy}$ and $p_z$ are precisely the values of the pressures found before by introducing the chemical potential.  

Notice that the calculation of  $p_{xy}$ and $p_z$ using the Brown-York tensor depends on the behavior of the geometry as we increase the holographic coordinate $r$ and approach the boundary.  On the contrary, the calculation of the pressures based on $\Phi$ is determined by the behavior of the geometry as we vary the flavor charge $Q_f$. The agreement of the results found by these two methods is a non-trivial consistency check of our gravity dual. 

\section{Effective action in 4d}
\label{4d_action}

In order to apply the full machinery of the holographic duality to our system it is quite convenient to integrate the action over the internal manifold and convert our problem into a system of low dimensional gravity.  There are two possible approaches to carry out this reduction. First of all, we could consider the 
$x^3$ coordinate as internal and reduce the system to a four-dimensional system in the coordinates 
$(t, x^1, x^2, r)$. This is the point of view we will adopt in this section.  This approach is very useful to study the dynamics of the system in the $(x^1, x^2)$ plane and, indeed, we will use the results of this section in our analysis of the hydrodynamics modes of section \ref{hydro}. Alternatively, we could include $x^3$ in our set of reduced coordinates and deal with a five-dimensional anisotropic problem. We will analyze this 5d reduction in the next section. 

The reduction of our problem to a low dimensional gravity system will allow us to implement a holographic renormalization procedure. We will be able to compute in this framework the VEV of the stress-energy tensor and to confirm the thermodynamic  results of section \ref{thermo_T_mu_nu}.  Moreover, in section \ref{hydro} we will study the fluctuations of the 4d fields and we  will obtain some hydrodynamic coefficients.

Let us consider the following  reduction ansatz  to four dimensions of the 10d metric:
\beq
ds_{10}^2\,=\,e^{{10\over 3}\,\gamma\,-\,\beta}\,g_{mn}dz^m\,dz^{n}\,+\,
e^{{10\over 3}\gamma\,+\,2\beta}\,(dx^3)^2\,+\,e^{-2(\gamma+\lambda)}\,ds^2_{{\mathbb C}{\mathbb P}^2}\,+\,
e^{2(4\lambda-\gamma)}\,(d\tau+A)^2\,\,,
\label{4d_metric_ansatz}
\eeq
where $g_{mn}=g_{mn}(z)$ is a 4d metric and the scalar fields $\gamma$, $\lambda$ and $\beta$ depend on the 4d coordinates $z^m\,=\,(t, x^1, x^2, r)$. In addition, in the reduced theory we have the dilaton field
$\phi=\phi(z)$. The action of this 4d gravity theory can be obtained from the one of type IIB supergravity. The details of this calculation are given in appendix \ref{Reduced_details}. The expression of this  effective action is:
\beq
S_{eff}\,=\,{V_5\,V_{x^3}\over 2\,\kappa_{10}^2}\,
\int d^4z\, \sqrt{-g_4}\,\,\Big[
R_4\,-\,{40\over 3}\,(\partial\gamma)^2\,-\,20\,(\partial\lambda)^2\,-\,{3\over 2}\,(\partial\beta)^2\,-\,
{1\over 2}\,(\partial\phi)^2\,-\,V\,\Big]\,\,,
\label{effective_4d_action}
\eeq
where $V$ is the following potential for the scalar fields $\phi$, $\gamma$, $\lambda$ and $\beta$:
\beq
V=4\,e^{{16\over 3}\,\gamma+12 \lambda-\beta}-24\,e^{{16\over 3}\gamma+2\lambda-\beta}+Q_f^2 \,e^{4\gamma+4\lambda-3\beta+\phi} +{Q_c^2\over 2}\,e^{{40\over 3}\gamma-\beta} + 6\,Q_f\,e^{{14\over 3}\gamma - 2\lambda - 2\beta\,+\,{\phi\over 2}}\,\,.
\label{4d_potential}
\eeq
In order to write the equations of motion of the reduced theory in a compact form, let us collect the four scalar fields in a single vector $\Psi$ with components:
\beq
\Psi=(\phi, \gamma, \lambda, \beta)\,\,.
\eeq
Moreover, we define a coefficient  $\alpha_{\Psi}$ which takes the following values for the different scalar fields:
\beq
(\alpha_{\phi}\,,\,\alpha_{\gamma}\,,\,\alpha_{\lambda}\,,\,\alpha_{\beta})\,=\,
(1\,,\,{3\over 80}\,,\,{1\over 40}\,,\,{1\over 3})\,\,.
\label{alpha_Psi}
\eeq
Then,   Einstein equations can be compactly written in terms of $\Psi$ as:
\beq
R_{mn}\,=\,\sum_{\Psi}\,{1\over 2\alpha_{\Psi}}\,
\partial_m\,\Psi\,\partial_n\,\Psi\,+\,{1\over 2}\,g_{mn}\,V\,\,.
\label{Einstein_eq_Psi}
\eeq
 Moreover, if we define the d'Alembertian of any scalar field $\Psi$ as:
\beq
 \Box \,\Psi \equiv {1\over \sqrt{-g_4}}\,\partial_m\,\Big( \sqrt{-g_4}\,g^{mn}\,\partial_n\,\Psi\Big)\,\,,
 \label{4d_Dalembertian}
\eeq
then, the equations for the scalar fields are:
\beq
\Box \,\Psi \,=\,\alpha_{\Psi}\,\partial_{\Psi}\,V\,\,.
\label{scalar_eom_4d_compact}
\eeq

Let us now write our black hole solution in terms of the 4d variables. The four-dimensional metric takes the diagonal form:
\beq
ds_4^2\,=\,-c_1^2(r)\,dt^2\,+\,c_2^2(r)[(dx^1)^2\,+\,(dx^2)^2]\,+\,c_3^2(r)\,dr^2\,\,.
\label{reduced_metric_cs}
\eeq
The actual values of the $c_i$ coefficients for our background are:
\bear
&&c_1^2(r)\,=\,\Big({9\over 8}\Big)^{3}\,\Big({4\,Q_f\over 3}\Big)^{{2\over 3}}\,R^2\,b(r)\,r^{{7\over 3}}
\,\,,\rc\rc
&&c_2^2(r)\,=\,\Big({9\over 8}\Big)^{3}\,\Big({4\,Q_f\over 3}\Big)^{{2\over 3}}\,R^2\,r^{{7\over 3}}\,=\,{c_1^2(r)\over b(r)}\,\,,\rc\rc
&&c_3^2(r)\,=\,\Big({9\over 8}\Big)^{3}\,\Big({4\,Q_f\over 3}\Big)^{{2\over 3}}\,
{R^6\over r^{{5\over 3}}\,b(r)}\,=\,{R^4\over r^4\,b^2(r)}\,c_1^2(r)\,\,,
\label{cs_4d_metric}
\eear
where $b(r)$ is the blackening factor defined in (\ref{blackening_factor_r}). 
Moreover, in our geometry the different scalars take the values:
\bear
&&e^{\phi}\,=\,\Big({3\over 4\, Q_f}\Big)^{{2\over 3}}\,r^{{2\over 3}}\,\,,
\qquad\qquad e^{\gamma}\,=\,\Big({8\over 9}\Big)^{{3\over 5}}\,\,{1\over R}\,\,,\rc\rc
&&e^{\lambda}\,=\,\Big({9\over 8}\Big)^{{1\over 10}}\,\,,
\qquad\qquad  e^{\beta}\,=\,{9\over 8}\,
\Big({4\,Q_f\over 3}\Big)^{{2\over 3}}\,\,R^{{2\over 3}}\,\,r^{{1\over 3}}\,\,.
\label{4d_scalars_solution}
\eear
One can easily verify that these metric and scalar fields solve (\ref{Einstein_eq_Psi}) and
(\ref{scalar_eom_4d_compact}). 

Let us have a closer look at the $4d$ metric we obtained. Plugging the $c_i(r)$ functions (\ref{cs_4d_metric}) into (\ref{reduced_metric_cs}), we get:
\beq
ds_4^2\,\sim\,r^{{7\over 3}}\,
\Big[-b(r)\,dt^2\,+\,(dx)^1+(dx^2)^2\,+\,R^4\,{dr^2\over b(r)\,r^4}\Big]\,\,.
\eeq
It is easy to check that this metric is equivalent to the one obtained when the  10d geometry of the D2-brane is reduced to 4d (change to the new radial coordinate $\rho=r^{{3\over 2}}$ and compare with the reduced metric written in \cite{Mas:2007ng}). Another way of reaching the same conclusion is by noting that under a scale transformation of the type:
\beq
t\,\to\,\lambda\,t\,,\qquad
x^{1,2}\,\to\,\lambda\,x^{1,2}\,,\qquad
r\,\to\,r/\lambda\,\,,
\eeq
the  zero-temperature metric changes homogeneously as:
\beq
ds_4^2\,\to\,\lambda^{-{1\over 3}}\,ds_4^2\,\,.
\eeq
This behavior corresponds to a hyperscaling violation of the type $ds^2_4\,\to\,\lambda^{\theta}\,ds^2_4$, with hyperscaling violation exponent $\theta=-{1\over 3}$ which, as shown in \cite{Dong:2012se}, is the $\theta$ exponent corresponding to a D2-brane. However, our 4d theory has more scalars than the reduced theory of a D2-brane and, therefore, even if the metrics are equal, both  problems are not equivalent in principle.

\subsection{Stress-energy tensor}

We now compute the VEV of the stress-energy tensor in this dimensionally reduced gravity theory. First of all we need to renormalize holographically the on-shell action by adding boundary terms. Besides the standard Gibbons-Hawking term, we will add a counterterm constructed with the superpotential  for the potential $V$ written in (\ref{4d_potential}) \cite{Batrachenko:2004fd}.  This superpotential will be denoted by $W_{4d}$ and must satisfy:
\beq
V\,=\,{1\over 2}\,\Big[{3\over 80}\,\big(\partial_{\gamma}\,W_{4d}\big)^2\,+\,{1\over 40}\,
\big(\partial_{\lambda}\,W_{4d}\big)^2\,+\,{1\over 3}\,\big(\partial_{\beta}\,W_{4d}\big)^2\,+\,
\big(\partial_{\phi}\,W_{4d}\big)^2
 \Big]\,-\,
{3\over 8}\,W_{4d}^2\,\,.
\label{V_W4d}
\eeq
It can be readily checked that the function:
\beq
W_{4d}\,=\,-6\,e^{{8\over 3}\,\gamma\,-\,4\lambda\,-\,{\beta\over 2}}\,-\,4
\,e^{{8\over 3}\,\gamma\,+\,6\lambda\,-\,{\beta\over 2}}\,+\,Q_c\,e^{{20\over 3}\gamma\,-\,{\beta\over 2}}\,+\,
2\,Q_f\,e^{2\gamma+2\lambda+{\phi\over 2}\,-\,{3\beta\over 2}}\,\,,
\label{W4d}
\eeq
solves (\ref{V_W4d}). Moreover, one can verify that $W_{4d}$ gives rise to the BPS equations satisfied by the zero temperature supersymmetric solution of \cite{Conde:2016hbg}. 

In terms of $W_{4d}$ the boundary action takes the form:
\beq
S_{boundary}\,=\,{V_5\,V_{x^3}\over 2\,\kappa_{10}^2}\,
\int_{r\to\infty}\,d^3 x\,\sqrt{\gamma}\,\Big(2\,K\,+\,W_{4d}\Big)\,\,,
\label{S_boundary_4d}
\eeq
where $\gamma$ is the determinant of the induced metric on constant-$r$ slices and $K=K^{\mu}_{\,\,\,\mu}$ is the trace of the extrinsic curvature of these slices. One can check that, after diving by the infinite volume $V_3$ of the $2+1$ dimensional Minkowski spacetime,  the sum of the actions (\ref{effective_4d_action}) and (\ref{S_boundary_4d}) evaluated on-shell is finite.  We get:
\beq
{S_{renormalized}\over V_3\,V_{x^3}}
\,=\,{S_{eff, on-shell}
\,+\,S_{boundary, on-shell}\over V_3\,V_{x^3}}\,=\,{3\over 7}\,\beta_s\,
Q_f^{{2\over 3}}\,Q_c^{{5\over 3}}\,T^{{10\over 3}}\,
\label{renormalized_onshell_4d_action}
\eeq
where $\beta_s$ is the constant defined in (\ref{beta_s}).  To obtain (\ref{renormalized_onshell_4d_action}) we have integrated from $r=r_h$ to $r=\infty$. Notice that $S_{renormalized}$ is equal, as it should, to minus the free energy density $f$ (compare with (\ref{free_energy_density})). The minus sign in this relation is due to the fact that we are working in Minkowski signature.

By taking the functional derivative of the on-shell renormalized action with respect to the boundary metric we obtain the expectation value of the field theory stress-energy tensor:
\beq
\langle  T^{\mu}_{\,\,\,\nu}\rangle
\,=\,{V_5\,V_{x^3}\over 2\,\kappa_{10}^2}\,
\sqrt{\gamma}\,\Big[-2\,K^{\mu}_{\,\,\,\nu}\,+\,\delta^{\mu}_{\,\,\,\nu}\,\big(2\,K\,+\,W_{4d}
\big)\Big]_{r\to\infty}\,\,.
\label{VEV_T_4d}
\eeq
Evaluating the right-hand side of (\ref{VEV_T_4d}) for our solution, we get:
\beq
\langle T^{\mu}_{\,\,\,\,\nu}\rangle\,=\,{\rm diag}\big(-\epsilon\,,\, p_{xy}\,,\, p_{xy}\,\big)\,\,,
\eeq
where $\epsilon$ is the ADM energy density  (\ref{energy_density}) and $p_{xy}$ is the pressure in the $xy$ plane written in (\ref{ps_Qs_epsilon}). 

\section{Effective action in 5d}
\label{5d_action}

Let us now reduce our system to five dimensions, namely those corresponding to the coordinates $z^m\,=\,(t, x^1, x^2, x^3, r)$. In principle this reduction would allow us to study the inter-layer properties and could be used to analyze the consequences of the anisotropy of the model. In this section we will use this 5d formalism to compute the complete stress-energy tensor and to establish the holographic dictionary for the D5-brane chemical potential.

Let us adopt the  following reduction ansatz for the metric will be:
\beq
ds_{10}^2\,=\,e^{{10\over 3}\,\gamma}\,g_{pq}\,dz^p\,dz^{q}\,+\,e^{-2(\gamma+\lambda)}\,ds^2_{{\mathbb C}{\mathbb P}^2}\,+\,
e^{2(4\lambda-\gamma)}\,(d\tau+A)^2\,\,,
\label{10d_5d_metric_ansatz}
\eeq
where $g_{pq}$ is a 5d metric and the scalar fields $\gamma$ and $\lambda$ depend on the 5d coordinates $z^m$.  It is important to notice that the RR three form $F_3$ for our solution has a leg in $x^3$, as well as two legs in the internal space (see (\ref{F3_ansatz})). Therefore, when it is reduced to 5d it gives rise to a one-form ${\cal F}_1$, which we will represent in terms of a scalar potential ${\cal V}$ as:
\beq
{\cal F}_1\,=\,d\,{\cal V}\,\,.
\label{F1-V}
\eeq
Moreover, our D5-branes are codimension-one objects (extended along the hypersurface $x^3={\rm constant}$ and smeared over $x^3$). The corresponding DBI action contains the determinant of the induced metric on this 4d surface, which we will denote by $\hat g_4$, integrated over $x_3$ to take into account the smearing. In addition to the metric and ${\cal V}$, the 5d theory has three scalar fields ($\gamma$, $\lambda$ and the dilaton $\phi$). 
The total effective action is worked out in appendix \ref{Reduced_details} and takes the form:
\bear
&&S_{eff}\,=\,{V_5\over 2\,\kappa_{10}^2}\,
\int d^5z\, \sqrt{-g_5}\,\,\Big[
R_5\,-\,{40\over 3}\,(\partial\gamma)^2\,-\,20\,(\partial\lambda)^2\,-\,
{1\over 2}\,(\partial\phi)^2\,-\,
{1\over 2}\,e^{4\gamma+4\lambda+\phi}\,(\partial {\cal V})^2\,-\,
U\,\Big]\,-\rc\rc
&&\quad\qquad\qquad\qquad\qquad\qquad
-\,{V_5\over 2\,\kappa_{10}^2}\,\int d^5z\, \sqrt{-\hat g_4}\big[\,6\,Q_f\,e^{{14\over 3}\,\gamma-2\lambda+{\phi\over 2}}\,\big]\,\,,
\label{effective_5d_action}
\eear
where $U$ is the potential:
\beq
U=4\,e^{{16\over 3}\,\gamma+12 \lambda}-24\,e^{{16\over 3}\gamma+2\lambda} +{Q_c^2\over 2}\,e^{{40\over 3}\gamma} \,\,.
\label{5d_potential}
\eeq
For our D3-D5 black hole solution the 5d  metric takes the form:
\beq
ds_5^2\,=\,-d_1^2(r)\,dt^2\,+\,d_2^2(r)\,\big[ (dx^1)^2\,+\,(dx^2)^2\big]\,+\,
d_3^2(r)\,(dx^3)^2\,+\,d_r^2(r)\,(dr)^2\,\,,
\label{5d_metric_ansatz}
\eeq
where the different $d$ functions are given by:
\bear
&&d_1^2(r)\,=\,\Big({9\over 8}\Big)^2\,r^2\,R^{{4\over 3}}\,b(r)\,\,,\rc\rc
&&d_2^2(r)\,=\,\Big({9\over 8}\Big)^2\,r^2\,R^{{4\over 3}}\,\,,\rc\rc
&&d_3^2(r)\,=\,\Big({9\over 8}\Big)^2\,\Big({4\,Q_f\over 3}\Big)^{{4\over 3}}\,R^{{4\over 3}}\,r^{{2\over 3}}\,\,,
\rc\rc
&&d_r^2(r)\,=\,\Big({9\over 8}\Big)^2\,{R^{{16\over 3}}\over r^2\,b(r)}\,\,.
\label{d_5dmetric}
\eear
In (\ref{d_5dmetric}) the function $b(r)$ is the blackening function (\ref{blackening_factor_r}). Moreover, the 
 scalar fields corresponding to the D3-D5 black hole are:
\beq
e^{\phi}\,=\,\Big({3\over 4\, Q_f}\Big)^{{2\over 3}}\,r^{{2\over 3}}\,\,,
\qquad\qquad e^{\gamma}\,=\,\Big({8\over 9}\Big)^{{3\over 5}}\,\,{1\over R}\,\,,
\qquad\qquad
e^{\lambda}\,=\,\Big({9\over 8}\Big)^{{1\over 10}}\,\,.
\qquad
\label{5d_scalars_solution}
\eeq
Notice that they are the same as in (\ref{4d_scalars_solution}). The function ${\cal V}$ is given by:
\beq
{\cal V}\,=\,\sqrt{2}\,Q_f\,x^3\,\,.
\label{cal_V_5d}
\eeq
It can be easily checked that the metric written in (\ref{5d_metric_ansatz}) and (\ref{d_5dmetric}), together  with the scalars written in (\ref{5d_scalars_solution}) and  the function ${\cal V}$ written in (\ref{cal_V_5d}), satisfy the equations of motion derived from the action (\ref{effective_5d_action}) (these equations have been explicitly written in appendix \ref{Reduced_details}). 

It is also interesting to relate these fields to the ones corresponding to the 4d approach for our solution. 
The 5d to 4d reduction is analyzed in appendix \ref{Reduced_details} (section \ref{5d_to_4d}).  As mentioned above, the scalars $(\phi, \gamma, \lambda)$ take the same values in 4d and 5d. Moreover, the 4d scalar 
$\beta$ is related to $d_3$ as:
\beq
e^{\beta}\,=\,d_3\,\,,
\label{beta_d3}
\eeq
while the functions $c_1$, $c_2$ and $c_3$ of the 4d metric are related to the $d$ functions as:
\beq
c_1^2\,=\,d_3\,d_1^2\,\,,
\qquad\qquad
c_2^2\,=\,d_3\,d_2^2\,\,,
\qquad\qquad
c_3^2\,=\,d_3\,d_r^2\,\,.
\label{cs_ds}
\eeq

\subsection{Stress-energy tensor}

Let us now construct boundary counterterms which regularize the on-shell effective action and allow to implement the holographic renormalization formalism and compute the VEV of the stress-energy  tensor.  First of all we obtain a superpotential $W_{5d}$ for the potential $U$  written in (\ref{5d_potential}). This superpotential must satisfy the equation:
\beq
U\,=\,{1\over 2}\,\Big[{3\over 80}\,\big(\partial_{\gamma}\,W_{5d}\big)^2
\,+\,{1\over 40}\,\big(\partial_{\gamma}\,W_{5d}\big)^2\,+\,
\big(\partial_{\phi}\,W_{5d}\big)^2\Big]\,-\,{1\over 3}\,W_{5d}^2\,\,,
\eeq
which is solved by the function:
\beq
W_{5d}\,=\,-6\,e^{{8\gamma\over 3}-4\lambda}\,-\,4\,
e^{{8\gamma\over 3}+6\lambda}\,+\,Q_c\,e^{{20\over 3}\gamma}\,\,.
\label{W5d}
\eeq
Notice that the three terms on the right-hand side of (\ref{W5d}) are in one-to-one correspondence with the terms in the 4d superpotential $W_{4d}$ which  do not contain $Q_f$ (see (\ref{W4d})).  Let us next define a new function $W_{flavor}$, related to the last term in (\ref{W4d}), as:
\beq
W_{flavor}\,=\,2\,Q_f\,e^{2\lambda+2\gamma+{\phi\over 2}}\,\,.
\label{W_flavor}
\eeq
The counterterms needed to renormalize the action (\ref{effective_5d_action}) will have the same structure as $S_{eff}$. First of all, we will have a 5d part, containing the metric $\gamma_{ab}$ induced on constant $r$ slices, as well as the Gibbons-Hawking term and the 5d superpotential (\ref{W5d}).  In addition, we will have a 4d part corresponding to the smeared sources, which contains the determinant of the metric $\hat\gamma_{ab}$ induced on constant $r$ and constant $x^3$ slices. We construct this term by using the flavor function defined in  (\ref{W_flavor}). The total boundary action is:
\beq
S_{boundary}\,=\,{V_5\over 2\,\kappa_{10}^2}\,
\int_{r\to\infty}\,d^4 x\,\sqrt{\gamma}\,\Big(2\,K\,+\,W_{5d}\Big)\,+\,
{V_5\over 2\,\kappa_{10}^2}\,\int_{r\to\infty}\,d^4 x\,\sqrt{\hat \gamma}\,
W_{flavor}\,\,.
\label{S_bd_5d}
\eeq
One can check that the addition of $ S_{boundary}$ makes the total on-shell action (divided by $V_3 V_{x^3}$)
 finite. Actually, one has:
\beq
{S_{renormalized}\over V_3 V_{x^3}} \,=\,{S_{eff, on-shell}\,+\,S_{boundary, on-shell}\over V_3 V_{x^3}} 
\,=\,
\,{3\over 7}\,\beta_s\,
Q_f^{{2\over 3}}\,Q_c^{{5\over 3}}\,T^{{10\over 3}}\,\,.
\eeq
Notice that $S_{renormalized}/V_3 V_{x^3}$ coincides with minus the free energy  density $f$ in the ten-dimensional approach (see  (\ref{free_energy_density})), as it should. 

The VEV of the stress-energy tensor of the dual theory can be obtained by taking the functional derivative of 
$S_{renormalized}$ with respect to the boundary metric. As a result of this calculation we get 
contributions from the two types of terms in (\ref{S_bd_5d}):
\beq
\langle  T^{\mu}_{\,\,\,\nu}\rangle
\,=\,{V_5\over 2\,\kappa_{10}^2}\,
\sqrt{\gamma}\,\Big[-2\,K^{\mu}_{\,\,\,\nu}\,+\,\delta^{\mu}_{\,\,\,\nu}\,\big(2\,K\,+\,W_{5d}
\big)\Big]_{r\to\infty}\,+\,
\langle  T^{\mu}_{\,\,\,\nu}\rangle_{flavor}\,\,,
\eeq
where $\langle  T^{\mu}_{\,\,\,\nu}\rangle_{flavor}$ is only non-vanishing if both indices $\mu$ and $\nu$ take values $0$, $1$,  $2$ and, in this case, is given by:
\beq
\langle  T^{\mu}_{\,\,\,\nu}\rangle_{flavor}\,=\,Q_f\,
{V_5\over \kappa_{10}^2}\,\sqrt{\hat\gamma}\, 
e^{2\lambda+2\gamma+{\phi\over 2}}\,\delta^{\mu}_{\,\,\,\nu}\Big|_{r\to \infty}\,\,,
\qquad\qquad\qquad
\mu,\nu=0,1,2\,\,.
\label{T_mu_nu_flavor}
\eeq
One can easily verify that $\langle  T^{\mu}_{\,\,\,\nu}\rangle$ is given by the same expression as in the 
10d analysis, namely by (\ref{T_values_10d}) with $\epsilon$, $p_{xy}$ and $p_z$ equal to the values written in  (\ref{energy_density}) and (\ref{ps_Qs_epsilon}).

\subsection{Holographic dictionary}

Clearly, the contribution (\ref{T_mu_nu_flavor})  is essential to reproduce the different values of the two pressures $p_{xy}$  and $p_z$, \ie\ to correctly represent the anisotropic behavior of the model. As argued in section \ref{thermodynamics}, this anisotropy is characterized by the D5-brane chemical potential $\Phi$.  It is therefore very important to find a dictionary allowing us to read the value of $\Phi$ from the value of some supergravity field at the UV boundary. This is the purpose of this subsection.

 In our holographic setup $\Phi$ should be related to the value of the potential under which the D5-branes are electrically charged. Notice that the D5-branes in our reduced theory extend along $x^0\,x^1\,x^2$ and  are smeared along $x^3$. Therefore, we expect $\Phi$ to be extracted from  the components of a three-form ${\cal C}_3$ along $x^0\,x^1\,x^2$. One can find 
${\cal C}_3$  by the following argument. First of all, we write the equation of motion of ${\cal F}_1$ (eq. (\ref{V_eom})) as:
\beq
d\Big(e^{4\gamma+4\lambda+\phi}\,*\,{\cal F}_1\Big)\,=\,0\,\,,
\label{eom_F1_forms}
\eeq
where $*$ denotes the Hodge dual of the 5d theory. Next, we interpret (\ref{eom_F1_forms})  as a Bianchi identity, \ie\ as the closure of the four-form $ {\cal F}_4$ defined as:
\beq
{\cal F}_4\,=\,e^{4\gamma+4\lambda+\phi}\,*\,{\cal F}_1\,\,.
\label{def_F4}
\eeq
It follows that ${\cal F}_4$ can be represented in terms of a three-form ${\cal C}_3$:
\beq
{\cal F}_4\,=\,d\,{\cal C}_3\,\,,
\label{def_C3}
\eeq
and we will soon verify that ${\cal C}_3$ is the three-form we are seeking. To check this statement we will find the form for our solution. First of all we notice that:
\beq
*\,{\cal F}_1\,=\,\sqrt{2}\,Q_f\,{d_1\,d_2^2\,d_r\over d_3}\,dx^0\wedge dx^1\wedge dx^2\wedge  dr\,\,.
\label{*F1_sol}
\eeq
From (\ref{*F1_sol}) we can readily verify that ${\cal C}_3$ can be taken as:
\beq
{\cal C}_3\,=\,{\cal A}\,\big(Q_f^{-{1\over 3}}\,r^{{10\over 3}}\,+\,{\cal C}\big)\,
dx^0\wedge dx^1\wedge dx^2\,\,,
\label{C3_sol}
\eeq
where ${\cal A}$ is a known numerical constant (independent of $Q_f$ and $Q_c$) and ${\cal C}$ is another constant which we will fix by requiring regularity at the horizon or, equivalently by demanding the vanishing of ${\cal C}_3$ at 
$r=r_h$. This condition leads to the following value of ${\cal C}$:
\beq
{\cal C}\,=\,-Q_f^{-{1\over 3}}\,r_h^{{10\over 3}}\,\,.
\label{cal_C_sol_rh}
\eeq
Taking into account that $r_h^{{10\over 3}}\propto Q_c^{{5\over 3}}\,T^{{10\over 3}}$, we find:
\beq
{\cal C}\,\propto\,Q_c^{{5\over 3}}\,Q_f^{-{1\over 3}}\,T^{{10\over 3}}\,\,.
\label{cal_C_sol_Qs}
\eeq
By comparing (\ref{cal_C_sol_Qs}) and (\ref{Phi_Qs_beta}) we conclude that the chemical potential $\Phi$ and the constant ${\cal C}$ are proportional:
\beq
\Phi\,\propto {\cal C}\,\,.
\eeq
Notice also that ${\cal C}$ is (proportional to)  the subleading term in the expansion of the $x^0x^1x^2$ component of ${\cal C}_3$ near the boundary. This identification of $\Phi$ is similar to the one obtained in \cite{Mateos:2011tv} for the case of an anisotropic background generated by D7-branes.  The fact that is the subleading term that is being identified with $\Phi$, and not the leading term as in other holographic setups, can be traced back to the Hodge duality that we are doing when passing from ${\cal F}_1$ to ${\cal F}_4$.

\section{Fluctuations and hydrodynamics}
\label{hydro}

We will now explore the hydrodynamic properties of our system. In particular we will compute the transport coefficients for perturbations propagating along the  $x^1x^2$ plane.  The purpose of this calculation is to characterize the effects of flavors, and of the corresponding induced anisotropy, on the transport properties of our system.  As already mentioned in the introduction, our main result is that the transport coefficients in the $x^1 x^2$ plane are the same as those of a D2-brane. This result confirms the conclusions of our static thermodynamic analysis and implies that, in our model, the dynamics of the excitations within a layer is governed by an effective strongly coupled super Yang-Mills theory in 2+1 dimensions.

Following the standard  procedure \cite{Kovtun:2005ev},  we have to study the fluctuations of the  4d metric and scalar fields around their background values 
(\ref{cs_4d_metric}) and (\ref{4d_scalars_solution}). In order to do this, we will perform the following substitution in the equations of motion:
\beq
g_{mn}\to g_{mn}+h_{mn}\,\,,
\qquad\qquad
\Psi\to \Psi \,+\,\delta \Psi\,\,,
\eeq
for $\Psi=(\phi, \gamma, \lambda, \beta)$ and we will keep only the first-order terms in $h_{mn}$ and 
$\delta\Psi$. Moreover, we will work in the radial gauge for the metric,  in which:
\beq
h_{mr}=0\,\,,
\qquad\qquad
(m=t, x^1, x^2, r)\,\,.
\label{radial_gauge}
\eeq
Let us start by computing the variation of the scalar equation (\ref{scalar_eom_4d_compact}). One can easily check that, at first order, we have:
\beq
\delta\,\Box \Psi\,=\,\Box \delta\Psi\,+\,{1\over 2}\,g^{mn}\,\partial_m\Psi\,\partial_n (h^p_{\,\,p})\,-\,
{1\over \sqrt{-g_4}}\,\partial_m\Big( \sqrt{-g_4}\,h^{mn}\,\partial_n\Psi\Big)\,\,.
\label{delta_box_Psi}
\eeq
The last  term in (\ref{delta_box_Psi}) is always zero in the radial gauge when the scalar fields of the background only depend on the radial variable. For a metric of the type (\ref{reduced_metric_cs}), $\Box \delta \Psi$  becomes:
\beq
\Box \delta \Psi={1\over c_3^2}\,\Big[
\partial_r^2\,(\delta \Psi)+\partial_r\log\Big({c_1\,c_2^2\over c_3}\Big)\,\partial_r\,(\delta \Psi)\Big]
-{\partial_t^2\,(\delta \Psi)\,\over c_1^2}\,+\,
{\partial_{x^1}^2\,(\delta \Psi)+\partial_{x^2}^2\,(\delta \Psi)\over c_2^2}\,\,,
\eeq
and the first-order equation for $\delta \Psi$ is:
\bear
&&\partial_r^2\,(\delta \Psi)+\partial_r\log\Big({c_1\,c_2^2\over c_3}\Big)\,\partial_r\,(\delta \Psi)
-{c_3^2\over c_1^2}\,\partial_t^2\,(\delta \Psi)\,\,+\,
{c_3^2\over c_2^2}\,\Big(\partial_{x^1}^2\,(\delta \Psi)+\partial_{x^2}^2\,(\delta \Psi)\Big)\,+\,\rc\rc
&&\qquad\qquad
+{\partial_r\Psi\over 2}\,\partial_r\,\Big({h_{x^1 x^1}+h_{x^2 x^2}\over c_2^2}\,-\,{h_{tt}\over c_1^2}\Big)\,=\,c_3^2\,\
\alpha_{\Psi}\,\delta[\partial_{\Psi} V]\,\,.
\label{fluct_eq_Psi}
\eear
The first-order variation of the Einstein equation (\ref{Einstein_eq_Psi}) is:
\beq
\delta\,R_{mn}\,=\,\sum_{\Psi}\,{1\over 2\alpha_{\Psi}}\,\Big(
\partial_m\,(\delta\Psi)\,\partial_n\,\Psi\,+\,
\partial_m\,\Psi\,\partial_n(\delta \Psi)\Big)\,
+\,{1\over 2 }\,h_{mn}\,V\,+\,{1\over 2 }\,g_{mn}\,\delta V\,\,,
\label{first_variation_Einstein}
\eeq
where $\delta\,R_{mn}$ can be written in terms of covariant derivatives of the metric perturbation $h_{mn}$ as:
\beq
\delta \,R_{mn}\,=\,{1\over 2}\,\Big[
D_p\,D_m\,h^{p}_{\,\,n}\,+\,
D_p\,D_n\,h^{p}_{\,\,m}\,-\,
D_p\,D^p\,h_{m n}\,-\,
D_m\,D^n\,h^p_{\,\, p}\Big]\,\,.
\label{delta_ricci}
\eeq
By plugging (\ref{delta_ricci}) into (\ref{first_variation_Einstein}), we arrive at the following equation for the metric fluctuations:
\bear
&&D_p\,D_m\,h^{p}_{\,\,n}\,+\,
D_p\,D_n\,h^{p}_{\,\,m}\,-\,
D_p\,D^p\,h_{m n}\,-\,
D_m\,D_n\,h^p_{\,\, p}\,=\,\rc\rc
&&\qquad
=\,\sum_{\Psi}\,{1\over \alpha_{\Psi}}\,\Big(
\partial_m\,(\delta\Psi)\,\partial_n\,\Psi\,+\,
\partial_m\,\Psi\,\partial_n(\delta \Psi)\Big)\,
+\,h_{mn}\,V\,+\,g_{mn}\,\delta V\,\,.
\qquad\qquad
\label{fluct_eq_h}
\eear

\subsection{The shear channel}
\label{shear_4d}

The fluctuation equations (\ref{fluct_eq_Psi}) and (\ref{fluct_eq_h}) are highly coupled. However, one can identify several consistent truncations in which only few fluctuations are non-zero.  Without loss of generality, let us consider a perturbation propagating along the $x^2$ direction.  The first of the consistent truncations that we will analyze is the so-called shear channel, in which only the metric fluctuations $h_{t\, x^1}$ and $h_{x^1\,x^2}$ are excited. Let us assume that these fluctuations  have frequency $\omega$ and momentum $q$ and, accordingly, let us parametrize them as:
\bear
&&h_{t\,x^1}\,=\,e^{-i(\omega\,t\,-\,q\,x^2)}\,c_2^2(r)\,H_{tx}(r)\,\,,\rc\rc
&&h_{x^1\,x^2}\,=\,e^{-i(\omega\,t\,-\,q\,x^2)}\,c_2^2(r)\,H_{xy}(r)\,\,,
\label{shear_ansatz}
\eear
where $c_2(r)$ is the function written in (\ref{4d_scalars_solution}) and has been included in the ansatz (\ref{shear_ansatz})  for convenience. The equations of motion of $H_{tx}$ and $H_{xy}$  are studied in detail in appendix \ref{hydro_details}. It turns out that they can be reduced to a single second-order differential equation for a gauge   invariant combination $X$, defined  as:
\beq
X\,\equiv\,q\,H_{tx}\,+\,\omega\,H_{xy}\,\,.
\label{X_def}
\eeq
The equation satisfied by $X$ is:
\beq
X''\,+\,{(10+3\,b(r))\omega^2\,-\,13\,b^2(r)\,q^2\over 
3\,b(r)\,r\,(\omega^2\,-\,b(r)\,q^2)}\,X'\,+\,
{R^4\over r^4\,b^2(r)}\,(\omega^2\,-\,b(r)\,q^2)\,X\,=\,0\,\,.
\label{eom_X}
\eeq
Let us now work in a new radial variable $x$, related to $r$ as:
\beq
x\,=\,\big[b(r)\big]^{{1\over 2}}\,\,.
\label{x_r_def}
\eeq
In this new variable the horizon is located at $x=0$, whereas the boundary is at $x=1$. 
We will consider the gauge-invariant combination $X$ as a function of $x$.  Moreover,  it is quite convenient to introduce the dimensionless momentum and frequency  $\hat q$ and $\hat \omega$, defined as:
\beq
\hat q\,=\,{q\over 2\pi\,T}\,\,,
\qquad\qquad
\hat\omega\,=\,{\omega\over 2\pi T}\,\,.
\label{hat_q_omega}
\eeq
Then, if the prime now denotes derivatives with respect to $x$, eq. (\ref{eom_X}) takes the form:
\beq
X''\,-\,{1\over x}\,{\hat q^2\,x^2\,+\,\hat\omega^2\over \hat q^2\,x^2\,-\,\hat\omega^2}\,X'\,-\,
{\hat q^2\,x^2\,-\,\hat\omega^2\over x^2(1-x^2)^{{7\over 5}}}\,X\,=\,0\,\,.
\label{X_eom_x}
\eeq
We want to solve (\ref{X_eom_x})  by imposing infalling boundary conditions at the horizon $x=0$, as well as Dirichlet  boundary conditions at the boundary $x=1$. These solutions only exist when the frequency $\omega$ and the momentum $q$ are related in a particular way, which determines the dispersion relation $\omega=\omega(q)$ of our modes. In the hydrodynamic regime the momentum $q$ is small and one can expand $\omega$ in a power series in $q$. In the shear channel we are studying this relation takes the form:
\beq
\omega\,=\,-i\,D_{\eta}\,q^2\,\big(1+\tau_s\,D_{\eta}\,q^2\big)\,\,,
\label{disp_rel_shear_q4}
\eeq
where  we are keeping terms up to quartic power of $q$. The dispersion relation (\ref{disp_rel_shear_q4}) depends on two transport coefficients $D_{\eta}$  and $\tau_s$, which we will calculate for our system in this section.  We will work in the  dimensionless variables  defined in (\ref{hat_q_omega}). Moreover, we define the rescaled coefficients $\hat D_{\eta}$ and 
$\hat\tau_s$ as:
\beq
\hat D_{\eta}\,=\,2\pi\,T\,D_{\eta}\,\,,
\qquad\qquad
\hat\tau_s\,=\,2\pi\,T\,\tau_s\,\,.
\eeq
In terms of the rescaled quantities, the dispersion relation (\ref{disp_rel_shear_q4}) takes the form:
\beq
\hat\omega\,=\,-i\,\hat D_{\eta}\,\hat q^2\,\big(1+\hat\tau_s\,\hat D_{\eta}\,\hat q^2\big)\,\,.
\eeq
The coefficient $\hat D_{\eta}$ determines the ratio of the  shear viscosity $\eta$ to the entropy density $s$,  namely: 
\beq
{\eta\over s}\,=\,{\hat D_{\eta}\over 2\pi}\,\,.
\eeq
Below we will  find that, for our system,  $\hat D_{\eta}=1/2$, which is equivalent to  having $\eta/s=1/(4\pi)$.  In what follows we  compute $\tau_s$ explicitly for our system and it turns out that $\tau_s$ is the same as the one found in \cite{Kapusta:2008ng} for the geometry of the D2-brane.

Let us come back to the integration of the differential equation (\ref{X_eom_x}). 
In order to impose infalling boundary conditions at the horizon $x=0$, we will adopt the ansatz:
\beq
X(x)\,=\,x^{-i\hat \omega}\,S(x)\,\,,
\eeq
where $S(x)$ must be regular at $x=0$. Let us expand $S(x)$ in powers of $\hat q$ as:
\beq
S(x)\,=\,S_0(x)\,+\,\hat q^2\,S_2(x)\,+\,\cdots\,\,.
\label{S_q_expansion}
\eeq
Plugging the expansions (\ref{S_q_expansion})  and (\ref{disp_rel_shear_q4}) into (\ref{X_eom_x}) and separating the different orders in $\hat q$, we get the following system of equations:
\bear
&&S_0''\,-\,{1\over x}\,S_0'\,=\,0\,\,,\rc\rc
&&S_2''\,-\,{1\over x}\,S_2'\,=\,\Bigg({1\over (1-x^2)^{{7\over 5}}}\,-\,{2\hat D_{\eta}\over x^2}\Bigg)\,S_0\,+\,
{2\hat D_{\eta}\over x}\Big(1-{\hat D_{\eta}\over x^2}\Big)\,S_0'\,\,.
\label{S0_S2_system}
\eear
We can also expand $S(x)$ in powers of $x$ near $x=0$:
\beq
S(x)\,=\,1\,+\,\sigma_2\,x^2\,+\,\sigma_4\,x^4\,+\,\cdots\,\,,
\eeq
where the coefficients $\sigma_2$ and $\sigma_4$ are easy to obtain by substituting this expansion into (\ref{X_eom_x}). They are given by:
\bear
&&
\sigma_2\,=\,{5i\,\hat q^2(2i+\hat\omega)\,-\,7i\,\hat\omega^3\over 20\,\hat\omega\,(i+\hat\omega)}\,\,,\rc\rc
&&\sigma_4\,=\,
{-25\,\hat q^4\,(4i+\hat\omega)\,+\,70\,\hat q^2\, \hat \omega (2i+\hat \omega)^2\,+\,
7\hat \omega^3(24-24 i\hat\omega-7\hat\omega^2)\over
800\,\hat\omega\,(i+\hat\omega)\,(2i+\hat \omega)}\,\,.
\eear
By expanding $\sigma_2$ and $\sigma_4$ in powers of $\hat q$ using the dispersion relation (\ref{disp_rel_shear_q4}), we arrive at the following expression of $S(x)$, valid for low $x$ and low $\hat q$:
\beq
S(x)\,=\,1\,-\,{x^2\over 2\hat D_{\eta}}\,+\,{\hat q^2\,x^2\over 80\,\hat D_{\eta}}\,
\Big(20\,\hat D_{\eta}\,(2\,\hat\tau_s\,-\,1)\,+\,(14\,\hat D_{\eta}\,-\,5)\,x^2\Big)\,+\,
{\cal O}(\hat q^3)\,\,.
\label{S_low_x_low_q}
\eeq
We will next compare (\ref{S_low_x_low_q}) with the result of integrating the system (\ref{S0_S2_system}) and expanding the result of this integration in powers of $x$ near $x=0$. The integration of the first equation in (\ref{S0_S2_system}) is straightforward and yields the result:
\beq
S_0(x)\,=\,A\,+\,B\,x^2\,\,,
\label{integral_S0}
\eeq
where $A$ and $B$ are integration constants.  By comparing (\ref{integral_S0}) with the first two terms in (\ref{S_low_x_low_q}) we conclude that $A=1$ and $B=-1/(2\hat D_{\eta})$ and, therefore, $S_0(x)$ is given by:
\beq
S_0(x)\,=\,1\,-\,{x^2\over 2\,\hat D_{\eta}}\,\,.
\eeq
By imposing the Dirichlet condition $S_0(x=1)=0$ at the boundary, we obtain  that, as already announced,  $\hat D_{\eta}$ must be:
\beq
\hat D_{\eta}\,=\,{1\over 2}\,\,,
\eeq
and $S_0$ takes the form:
\beq
S_0(x)\,=\,1\,-\,x^2\,\,.
\eeq
Using these values of $S_0(x)$ and $\hat D_{\eta}$ on the right-hand side of the second equation of the system (\ref{S0_S2_system}) we arrive at the equation:
\beq
S_2''\,-\,{1\over x}\,S_2'\,=\,{1\over (1-x^2)^{{2\over 5}}}\,-\,1\,\,,
\eeq
whose general solution is:
\beq
S_2(x)\,=\, C\,+\, (1+2D-2\log x)\,{x^2\over 4}\,-\,
{25\over 24}\,e^{{2\pi i\over 5}}\,x^{{6\over 5}}\,
F\Big(-{3\over 5}, {2\over 5}; {7\over 5}; {1\over x^2}\Big)\,\,.
\label{integral_S2}
\eeq
In (\ref{integral_S2}) $C$ and $D$ are integration constants which can be determined by expanding the result near $x\approx 0$ and comparing it with the terms proportional to $\hat q^2$ of (\ref{S_low_x_low_q}).  The expansion of (\ref{integral_S2})   near  $x\approx 0$ is:
\beq
S_2(x)\,=\,C\,+\,{5\over 12}\,+\,{x^2\over 2}\,
\Bigg[D\,+\,{1\over 2}\,\Big(\gamma-i\pi+\psi\Big({{2\over 5}}\Big)\Big)\Bigg]\,+\,
{x^4\over 20}\,+\,\cdots\,\,,
\eeq
where $\gamma\approx .577$ is the Euler-Mascheroni constant and $\psi(z)$ is the logarithmic derivative of the Euler gamma function $\Gamma(z)$.  This result coincides with (\ref{S_low_x_low_q}) if the constants $C$ and $D$ are:
\bear
&&C\,=\,-{5\over 12}\,\,,\rc\rc
&&D\,=\,-{1\over 2}\,\Big[1\,-\,2\,\hat\tau_s\,+\,
\gamma-i\pi+\psi\Big({{2\over 5}}\Big)\Big]\,\,.
\label{C_D_of_S2}
\eear
Substituting (\ref{C_D_of_S2}) in (\ref{integral_S2}) we get  the function $S_2(x)$, namely:
\beq
S_2(x)=-{5\over 12}+\Big(2\,\hat\tau_s-\gamma+i\pi-\psi\Big({{2\over 5}}\Big)-
2\log x\Big){x^2\over 4}-
{25\over 24}e^{{2\pi i\over 5}}\,x^{{6\over 5}}\,
F\Big(-{3\over 5}, {2\over 5}; {7\over 5}; {1\over x^2}\Big)\,\,,
\eeq
which only contains $\hat\tau_s$ as unknown parameter.  By imposing that $S_2(x=1)=0$, $\hat\tau_s$ is fixed to be:
\beq
\hat\tau_s\,=\,{1\over 2}\,\Big[\gamma+\psi\Big({{8\over 5}}\Big)\Big]\,\,.
\eeq
Equivalently, the unrescaled parameter is:
\beq
\tau_s\,=\,{1\over 4\,\pi\,T}\,\Big[\gamma+\psi\Big({{8\over 5}}\Big)\Big]\,\,.
\eeq
This value of $\tau_s$ coincides with the one found in the literature for the D2-brane \cite{Kapusta:2008ng}.

\subsection{The sound channel}

In the so-called sound channel, the following set of metric fluctuations, propagating along $x^2$,  are decoupled from the others:
\beq
(h_{tt}, h_{tx^2}, h_{x^1 x^1}, h_{x^2 x^2})\,\,,
\eeq
and are coupled to the fluctuations of the scalar fields. Let us parametrize these metric fluctuations as:
\bear
&&h_{tt}\,=\,e^{-i(\omega\,t\,-\,q\,x^2)}\,c_1^2(r)\,H_{tt}(r)\,\,,
\qquad\qquad\,\,\,\,
h_{tx^2}\,=\,e^{-i(\omega\,t\,-\,q\,x^2)}\,c_2^2(r)\,H_{ty}(r)\,\,,\qquad\qquad\rc\rc
&&h_{x^1 x^1}\,=\,e^{-i(\omega\,t\,-\,q\,x^2)}\,c_2^2(r)\,H_{xx}(r)\,\,,
\qquad\qquad
h_{x^2 x^2}\,=\,e^{-i(\omega\,t\,-\,q\,x^2)}\,c_2^2(r)\,H_{yy}(r)\,\,,\qquad\qquad
\label{metric_waves}
\eear
where $c_1(r)$ and $c_2(r)$ are the functions written in (\ref{cs_4d_metric}). 
Similarly, we represent the scalar fluctuations as:
\bear
&&\delta\phi\,=\,e^{-i(\omega\,t\,-\,q\,x^2)}\,\Phi(r)\,\,,\qquad\qquad\qquad
\delta\gamma\,=\,e^{-i(\omega\,t\,-\,q\,x^2)}\,\Gamma(r)\,\,,\qquad\qquad\rc\rc
&&\delta\lambda\,=\,e^{-i(\omega\,t\,-\,q\,x^2)}\,\Lambda(r)\,\,,\qquad\qquad\qquad
\delta\beta\,=\,e^{-i(\omega\,t\,-\,q\,x^2)}\,B(r)\,\,.\qquad\qquad
\label{scalar_wawes}
\eear
Let us now introduce a compact notation for the scalar fluctuations. We denote by $\hat\Psi(r)$  the radial part  of the fluctuation $\Psi=(\phi,\gamma,\lambda, \beta)$, namely:
\beq
\hat\Psi(r)\,=\,(\Phi(r), \Gamma(r), \Lambda(r), B(r))\,\,.
\eeq
Then, (\ref{scalar_wawes}) can be rewritten simply as:
\beq
\delta\Psi\,=\,e^{-i(\omega\,t\,-\,q\,x^2)}\hat\Psi(r)\,\,.
\label{scalar_wawes_compact}
\eeq
The full set of equations for the fields of (\ref{metric_waves}) and (\ref{scalar_wawes}) is written in appendix \ref{hydro_details}.  As usual, these equations are highly redundant due to the diffeomorphism gauge invariance. This redundancy can be reduced by defining new fields. Accordingly,  let us define new scalar fluctuation fields $Z_{\Phi}$, $Z_{\Gamma}$, $Z_{\Lambda}$ and $Z_{B}$, denoted collectively by $Z_{\hat\Psi}$, as the following combination of $\hat\Psi$ and 
$H_{xx}$:
\beq
Z_{\hat\Psi}\,=\,\hat\Psi\,-\,{\Psi'\over \partial_r\,\log c_2^2}\,\,H_{xx}\,\,.
\label{Z_Psi_def}
\eeq
As argued  in \cite{Benincasa:2005iv,Benincasa:2006ei,Mas:2007ng}, these are the gauge invariant combinations of the scalar fields  and the metric.    It is proved in appendix \ref{hydro_details} that the equations for the $Z$'s close among themselves (see the system (\ref{eoms_Zs})). Moreover, there is a particular combination $Z_S$ of these fields which can be decoupled from the other scalars. This combination is:
\beq
Z_{S}(r)\,\equiv\,3\,Z_{B}(r)\,+\,2\,Z_{\Phi}(r)\,\,.
\label{Z_S_def}
\eeq
The equation satisfied by $Z_{S}(r)$ has been written in (\ref{ZS_eom_explicit}). 
Following \cite{Kovtun:2005ev},  we  now define the gauge invariant metric fluctuation $Z_H$ as:
\beq
Z_H\,=\,H_{yy}\,+\,{2 q\over \omega}\,H_{ty}\,+\,
{q^2\over \omega^2}\,{c_1^2\over c_2^2}\,H_{tt}\,+\,
\Big({q^2\over \omega^2}\,{c_1^2\partial_r\log c_1\over c_2^2\partial_r\log c_2}\,-\,1\Big)\,H_{xx}\,\,.
\eeq
The equation satisfied by $Z_H$ has been written in (\ref{ZH_eom}). This equation shows that $Z_H$ is only coupled to $Z_S$.  Since $Z_S$ does not couple to any other scalar, we can start our analysis by finding $Z_S$ and then using this result in the equation for $Z_H$. It is shown in appendix \ref{hydro_details} that the only acceptable solution for $Z_S$ is the trivial one $Z_S=0$. Thus, we are left with a single equation for the gauge invariant metric fluctuation $Z_H$. Let us adopt for $Z_H$ an ansatz similar to the one used for the fluctuations in the shear channel, namely:
\beq
Z_H(r)\,=\,\big[ b(r)\big]^{-{i\hat\omega\over 2}}\,\,Y(r)\,\,.
\eeq
Furthermore, we will work in the $x$ variable defined in (\ref{x_r_def}). After some work one can verify that the equation satisfied by $Y(x)$ is:
\bear
&&Y''\,+\,
{\big[5-2(3+ 2i\hat \omega)x^2-10i\,\hat \omega\big]\hat q^2\,+\,
7(2i\,\hat\omega\,-\,1)\hat\omega^2\over
x\big[(5+2 x^2)\hat q^2\,-\,7\,\hat \omega^2\big]}\,\,Y'\,+\,\rc\rc
&&\qquad
+\,\Big[-{\hat q^2\over (1-x^2)^{{7\over 5}}}\,+\,{1\over x^2}\,
\Big({1\over (1-x^2)^{{7\over 5}}}\,-\,1\Big)\,\hat\omega^2\,+\,
{8(1+i\hat \omega)\,\hat q^2\over (5+2 x^2)\hat q^2\,-\,7\,\hat \omega^2}\Big]\,Y\,=\,0\,\,,
\qquad\qquad
\label{Y_eq}
\eear
where the primes denote derivative with respect to the new variable $x$.

We want to integrate the differential equations for $Y(x)$ in the hydrodynamic limit of low momentum.  We will impose infalling boundary conditions at the horizon for $Z_H(r)$  and we will demand that the fluctuations vanish at the boundary. The infalling boundary condition at the horizon $x=0$ is equivalent to the regularity of $Y(x)$ at this point.  These conditions would require a specific dispersion relation 
$\omega=\omega(q)$, which at low momentum can be expanded as:
\beq
\omega\,=\,v_s\, q\,-\,i\Gamma\, q^2\,+\,{\cal T} \, q^3\,\,,
\label{dispersion}
\eeq
where we have only kept terms up to third order in $q$. The coefficient $v_s$ of the linear term in 
(\ref{dispersion}) is the speed of sound and the quadratic coefficient $\Gamma$ is the attenuation which, 
in $p$ spatial dimensions, is related to the shear viscosity $\eta$ and the bulk viscosity $\zeta$ as:
\beq
\Gamma\,=\,{1\over T\,s}\,\Big[{p-1\over p}\,\eta\,+\,{\zeta\over 2}\Big]\,\,,
\eeq
where $s$ is the entropy density. In our $p=2$ case this expression becomes:
\beq
\Gamma\,=\,{1\over 2\,T\,s}\,(\eta\,+\,\zeta)\,\,.
\label{Gamma_eta_zeta}
\eeq
The cubic coefficient ${\cal T}$ is usually \cite{Buchel:2009hv}  parametrized as:
\beq
{\cal T}\,=\,{\Gamma\over v_s}\,\Big[v_s^2\,\tau_{eff}\,-\,{\Gamma\over 2}\Big]\,\,,
\eeq
where  $\tau_{eff}$ is an effective equilibration time which, in $p$ spatial dimensions, is related to the second-order transport coefficients $\tau_{\pi}$ and $\tau_{\Pi}$ of the Israel-Stewart theory as:
\beq
\tau_{eff}\,=\,{\tau_{\pi}\,+\,{p\over 2(p-1)}\,{\zeta\over \eta}\,\tau_{\Pi}\over
1+{p\over 2(p-1)}\,{\zeta\over \eta}}\,\,.
\eeq
In our $p=2$ model we have:
\beq
\tau_{eff}\,=\,{\tau_{\pi}\,+\,{\zeta\over \eta}\,\tau_{\Pi}\over
1+\,{\zeta\over \eta}}\,\,.
\eeq
In what follows it is quite convenient to work with  the dimensionless momentum and frequency 
$\hat q$ and $\hat \omega$ defined in (\ref{hat_q_omega}).  In terms of these rescaled quantities, the dispersion relation (\ref{dispersion}) takes the form:
\beq
\hat\omega\,=\,v_s\,\hat q\,-\,i\hat\Gamma\,\hat q^2\,+\,\hat {\cal T} \,\hat q^3\,\,,
\eeq
where $\hat\Gamma$ and $\hat{\cal T}$ are related to  $\Gamma$ and  ${\cal T}$  as:
\beq
\hat\Gamma\,=\,2\,\pi\,T\,\Gamma\,\,,
\qquad\qquad
\hat {\cal T}\,=\,(2\pi T)^2\,{\cal T}\,\,.
\label{rescaled_Gamma_T}
\eeq

Let us now analyze (\ref{Y_eq}) in the hydrodynamic approximation. We first expand $Y(x)$ in powers of $\hat q$ (up to second order) as:
\beq
Y(x)\,=\,Y_0(x)\,+\,i\hat q\,Y_1(x)\,+\,\hat q^2\,Y_2(x)\,\,.
\label{eta_q_expansion}
\eeq
By using (\ref{eta_q_expansion}) and (\ref{dispersion})  in (\ref{Y_eq}), one can  readily show that $Y_0(x)$ satisfies the equation:
\beq
Y_0''\,+\,{5-7\,v_s^2\,-\,6\,x^2\over x(5\,-\,7 v_s^2\,+\,2\, x^2)}\,Y_0'\,+\,
{8\over 5-7\,v_s^2\,+2\, x^2}\,Y_0\,=\,0\,\,,
\eeq
whose  general solution is:
\beq
Y_0(x)\,=\,C_1\,\Big(1+{2 x^2\over 7 v_s^2-5}\Big)\,+\,
C_2\,\Big(2(7v_s^2-5)\,+\,(2x^2-5+7v_s^2)\,\log x\Big)\,\,,
\eeq
where $C_1$ and $C_2$ are integration constants. 
Regularity at the horizon ($x=0$) requires that $C_2=0$. By imposing that 
\beq
Y_0(x=1)\,=\,0\,\,,
\eeq
we get the speed of sound $v_s$, namely:
\beq
v_s\,=\sqrt{{3\over 7}}\,\,,
\label{vs_sound_mode}
\eeq
which coincides with the value  we found in our static analysis for the propagation in the $x^1 x^2$ plane, \ie\ it is the same as the speed of sound propagating along the gauge theory directions of a D2-brane. 

Without loss  of generality we can take $C_1=1$ (or, equivalently, $Y_0(x=0)=1$) and, therefore, $Y_0(x)$ becomes:
\beq
Y_0(x)\,=\,1-x^2\,\,.
\label{Y_0}
\eeq
The  equation for $Y_1$ is:
\bear
&&Y_1''\,+\,{5-7\,v_s^2-6\, x^2\over x(5-7\,v_s^2+2\,x^2)}\,Y_1'\,+\,
{8\over 5-7\,v_s^2+2\,x^2}\,Y_1\,=\,\rc\rc
&&\qquad
={8 v_s\over 5-7\,v_s^2+2\,x^2}\,\Big[{14\,\hat\Gamma\over 5-7\,v_s^2+2\,x^2}\,-\,1\Big]\,Y_0\,+\,
{v_s\over x}\Big[2-{112 \,x^2\,\hat\Gamma\over (5-7\,v_s^2+2\,x^2)^2}\Big]\,Y_0'\,\,.
\qquad\qquad
\eear
Using the values of $Y_0$ and $v_s$ written in (\ref{Y_0}) and (\ref{vs_sound_mode}) this equation becomes:
\beq
Y_1''+{1-3 x^2\over x(1+x^2)}\,Y_1'\,+\,{4\over 1+x^2}\,Y_1\,=\,4 \sqrt{{3\over 7}}\,\,
{7\,\hat\Gamma-2\over 1+x^2}\,\,.
\label{eta_1_eq_simple}
\eeq
The general solution of (\ref{eta_1_eq_simple}) is:
\beq
Y_1(x)\,=\,\sqrt{{3\over 7}}\,(7\,\hat\Gamma-2)\,+\, C_1\,(1-x^2)\,+\,C_2\,
\big(4+(1-x^2)\log x^2\big)\,\,,
\label{eta_1_general_sol}
\eeq
where, again, $C_1$ and $C_2$ are integration constants.  The regularity requirement at the horizon $x=0$ implies that $C_2=0$. Moreover, the UV condition $Y_1(x=1)=0$ fixes the rescaled attenuation to be:
\beq
\hat\Gamma\,=\,{2\over 7}\,\,,
\eeq
which, according to (\ref{rescaled_Gamma_T}), is equivalent to the following value of $\Gamma$:
\beq
\Gamma\,=\,{1\over 7\pi \,T}\,\,.
\label{Gamma_value}
\eeq
Taking into account that $\eta/s=1/4\pi$, it follows from (\ref{Gamma_value}) and (\ref{Gamma_eta_zeta}) 
 that the ratio of the bulk and shear viscosities for our model is:
\beq
{\zeta\over \eta}\,=\,{1\over 7}\,\,,
\eeq
This value for $\zeta/\eta$ is exactly the same as the one corresponding to a D2-brane \cite{Mas:2007ng}, which saturates Buchel's bound\cite{Buchel:2007mf}:
\beq
{\zeta\over \eta}\,=\,2\Big({1\over 2}\,-\,v_s^2\Big)\,\,.
\eeq
Let us next look at the equation for $Y_2(x)$. Using the values of $v_s$ and $\hat \Gamma$ already determined, this equation reduces to:
\beq
Y_2''+{1-3 x^2\over x(1+x^2)}\,Y_2'\,+\,{4\over 1+x^2}\,Y_2\,=\,g(x)\,\,,
\label{Y_2_eq}
\eeq
where $g(x)$ is the following function:
\beq
g(x)\,=\,{1\over (1-x^2)^{{2\over 5}}}\,-\,{3\over 7}\,\Big(1\,-\,{1\over x^2}\,+\,
{1\over x^2 (1-x^2)^{{2\over 5}}}\Big)\,-\,{4\over 7}\,\,
{7\sqrt{21}\,\hat{\cal T}\,-\,4\over 1+x^2}\,\,.
\eeq
The homogeneous equation in (\ref{Y_2_eq}) is just the same as in (\ref{eta_1_eq_simple}). We already found two independent solutions in (\ref{eta_1_general_sol}), which we now denote by $y_1(x)$ and $y_2(x)$:
\beq
y_1(x)\,=\,1\,-\,x^2\,\,,
\qquad\qquad
y_2(x)\,=\,(1-x^2)\,\log x^2\,+\,4\,\,.
\eeq
Then, the general solution of (\ref{Y_2_eq}) can be written as:
\beq
Y_2(x)\,=\,D_1\,y_1(x)\,+\,D_2\,y_2(x)\,+\,y_p(x)\,\,,
\label{general_sol_eta2}
\eeq
where $D_1$ and $D_2$ are constants and  $y_p(x)$ is a particular solution of the full inhomogeneous equation. We will use the method of variation of constants to find $y_p(x)$. The result can be written as:
\beq
y_p(x)\,=\,y_2(x)\,\,\int dx\, {y_1(x)\,g(x)\over W(x)}\,-\,y_1(x)\,\,\int dx\, {y_2(x)\,g(x)\over W(x)}\,\,,
\label{particular_wronskian}
\eeq
where $W(x)$ is the Wronskian:
\beq
W(x)\,=\,y_1(x)\,y_2'(x)\,-\,y_1'(x)\,y_2(x)\,\,.
\eeq
Let us rewrite (\ref{particular_wronskian}) in a more convenient way following \cite{Springer:2009wj,Springer:2009tu}. First of all, we define $h(x)$ as the ratio between the two solutions of the homogeneous equation:
\beq
h(x)\,=\,{y_2(x)\over y_1(x)}\,\,.
\eeq
The Wronskian $W(x)$ is related to the derivative of $h(x)$ as:
\beq
W(x)\,=\,h'(x)\,y_1^2(x)\,\,,
\eeq
and, therefore, we can rewrite (\ref{particular_wronskian}) as:
\beq
y_p(x)\,=\,y_1(x)\,\Big[ h(x)\,\int dx {g(x)\over y_1(x)\,h'(x)}\,-\,\int dx {h(x)\,g(x)\over h'(x)\,y_1(x)}\Big]\,\,.
\eeq
After an integration by parts, this equation can be recast as:
\beq
y_p(x)\,=\,y_1(x)\,\int dx \,h'(x)\,\int^{x}\,{g(z)\over y_1(z)\,h'(z)}\,dz\,\,.
\label{y_p_integral}
\eeq
We now  impose  the regularity condition at the horizon $x=0$. Using the integral expression 
(\ref{y_p_integral})  one can show that  near $x\to 0$ the solution behaves as:
\beq
Y_2(x)\approx {\cal A}\,+\,{\cal B}\,\log x\,+\,\cdots\,\,
\label{eta_2_horizon}
\eeq
where ${\cal A}$ and ${\cal B}$ are constants and  the dots represent terms that vanish at $x=0$. Our regularity condition demands that the term with the logarithm be absent in (\ref{eta_2_horizon}).  Then, we require:
\beq
{\cal B}\,=\,0\,\,.
\eeq
This determines the constant  $D_2$ in (\ref{general_sol_eta2}) to be:
\beq
D_2\,=\,{1\over 14}\,\Big[ {3\over 2}\,\Big(\gamma\,-\,i\pi\,+\,\psi\Big({2\over 5}\Big)\Big)\,-\,1\Big]\,\,,
\eeq
where $\gamma=0.577$ is the Euler-Mascheroni constant and $\psi(x)=\Gamma'(x)/\Gamma(x)$ is the digamma function.  We next impose the UV  boundary condition at $x=1$: 
\beq
Y_2(x\to 1)\,=\,0\,\,,
\eeq
which determines the  value of $\hat{\cal T}$ as:
\beq
\hat{\cal T}\,=\,{1\over 7}\sqrt{{3\over 7}}\,\Big(1+\gamma+\psi\Big({8\over 5}\Big)\Big)\,\,.
\eeq
Numerically, $\hat{\cal T}\approx 0.1592$.  Using (\ref{rescaled_Gamma_T}) we find the following value of ${\cal T}$:
\beq
{\cal T}\,=\,{\sqrt{3}\over 28\sqrt{7}\,(\pi\,T)^2}\,
\Big(1+\gamma+\psi\Big({8\over 5}\Big)\Big)\,\,.
\eeq
Taking into account the value of $\Gamma$ we found (eq. (\ref{Gamma_value})),  this result corresponds to having an equilibration time $\tau_{eff}$ equal to:
\beq
\tau_{eff}\,=\,{1\over 4\pi\,T}\Big[{5\over 3}\,+\,\gamma\,+\psi\Big({8\over 5}\Big)\Big]\,\,,
\eeq
which again coincides with the one found for the geometry of a D2-brane \cite{Springer:2009wj,Springer:2009tu}. 
From this value of $\tau_{eff}$ we get  the following relation between the two  Israel-Stewart 
coefficients, namely:
\beq
7\,\tau_{\pi}\,+\,\tau_{\Pi}\,=\,{2\over \pi T}\,
\Big[{5\over 3}\,+\,\gamma\,+\,\psi\Big({8\over 5}\Big)\Big]\,\,.
\eeq

\section{Summary and conclusions}
\label{conclusions}

Let us summarize our main results. We have succeeded in generalizing the D3-D5 geometry of \cite{Conde:2016hbg} to include an event horizon. Our solution is analytic and simple and is the gravity dual of the defect theory introduced in \cite{DeWolfe:2001pq} at non-zero temperature in the approximation in which the massless flavors are smeared.  The geometry found is homogeneous but anisotropic in the gauge theory directions: it preserves translational invariance but breaks rotational symmetry.

We have studied the thermodynamics and hydrodynamics of the model. We have checked several thermodynamic relations and found that the results are consistent with the laws of  anisotropic thermodynamics.  We  also obtained dimensionally reduced  gravitational actions for our system in four and five dimensions. In both dimensionalities we managed to construct boundary terms to renormalize the on-shell action and find the stress-energy tensor. Moreover, we obtained the hydrodynamic transport coefficients (up to second order) for perturbations propagating in the $x^1x^2$ plane. These transport coefficients are exactly the same as those of the D2-brane, a result which is not obvious despite the 2+1 dimensionality of our defect theory. 

It follows from  our results  that the energy of our system scales with $Q_c$ and $Q_f$ as $Q_c^{{5\over 3}}\,Q_f^{{2\over 3}}$, which determines  the  dependence of  the effective number of degrees of freedom on the number of colors and flavors.  This type of dependence with $Q_c$ and $Q_f$ shows up in our thermodynamic results of section \ref{thermodynamics}, as well as in the dependence of the entanglement entropy $S_{\parallel}$ (see eq. (\ref{S_parallel})). The non-integer powers of  $Q_c$ and $Q_f$ in this scaling are reflecting  the strong coupling regime of the dynamics of the layers. The main result of our thermodynamic and hydrodynamic analysis is that this layer behavior can be reproduced by an effective D2-brane or, equivalently, by 2+1 super Yang-Mills in the strong coupling regime.

Let us discuss some possible extensions of our work. We could use our entanglement entropy results  for slabs of appendix \ref{WL-EE} to study the quantum correlations of the model. From the dependence of the entanglement entropy on the width of the slab it should be immediate to study the mutual information of two slabs and to analyze the possible phase transitions. Moreover,  we could also test our geometry with different probe branes, which would correspond to adding new degrees of freedom. One possibility would be adding D5-brane probes of the same type as the ones that originated the background and studying their thermodynamics as in \cite{Mateos:2007vn}. In this probe brane setup it is rather easy to add a baryonic chemical potential. Another possibility would be adding D7-branes extended along the four Minkowski directions, which would allow us to study the anisotropy of the model from a different point of view.

We have restricted our hydrodynamic study to modes propagating in the $x^1 x^2$ plane. It would be very interesting to extend this analysis to modes propagating along $x^3$ and to  explore the effects of anisotropy on the transport coefficients. To carry out this task we should make use of the 5d reduced action found in section  \ref{5d_action}.  However, this reduced model contains a codimension one object embedded in the fixed hypersurface  $x^3={\rm constant}$ (and smeared over $x^3$). The fluctuations of this action involving the $x^3$ direction are very difficult to treat and we could not find the analogue of the decoupled gauge invariant combinations of section \ref{hydro}. On general grounds we would expect to find the same speed of sound $v_z$ as in (\ref{speeds_of_sound}). In the shear channel we could violate the KSS bound, as it happens in other anisotropic models 
\cite{Rebhan:2011vd,Mamo:2012sy}. As a preliminary calculation one can consider the perturbation of the $x^1 x^3$ component of the metric and study the response function. By using the standard Kubo formalism in the holographic setup, we get (see, for example, \cite{Jain:2014vka}):
\beq
{\eta_{\perp}\over s}\,=\,{g_{x^1x^1}\over g_{x^3x^3}}\Bigg|_{r=r_h}\,\,.
\eeq
The value of the transverse viscosity $\eta_{\perp}$ obtained in this way satisfies $\eta_{\perp}/s\sim Q_c^{{2\over 3}}\,Q_f^{-{4\over 3}}\,T^{{4\over 3}}$, which certainly can be arbitrary small as $T\to 0$ and, therefore, violates the KSS bound at low temperatures. 

One important feature of our geometry is that it does not have a weak anisotropy limit and, in fact, it is non-analytic in $Q_f$ when $Q_f\to 0$. This is due to the fact that the flavors introduced are massless. It was shown in \cite{Conde:2016hbg} how to generalize the supersymmetric ($T=0$) solution to the case in which the flavors are massive. In this case the flavor branes do not reach the origin and there is a cavity around $r=0$ in which the D5-brane  charge is zero and the equations of motion are those of the unflavored system. The radius of the cavity is related to the mass of the quarks. The massive solutions found in \cite{Conde:2016hbg} interpolate between the unflavored metric in the IR and the massless flavored geometry in the UV. By sending the quark mass to infinity the size of the cavity increases and the geometry becomes $AdS_5\times {\mathbb S}^5$. This is quite natural from the point of view of field theory since in this infinite mass limit we are making  the flavors non dynamical. From the holographic point of view,  the quark mass is an external parameter which allows to modify the degree of anisotropy. It would be very interesting to generalize some of the results found here to this massive case and to explore the development of anisotropy and their effects on the physical observables. Work along these lines is in progress.

\vspace{2cm}
{\bf \large Acknowledgments}

We are grateful to Y. Bea, F. Bigazzi,  G. Itsios, N. Jokela, D. Musso, C. N\'u\~nez  and J. Tarr\'\i o for discussions and useful suggestions.  J. M. P. and A. V. R.   are funded by the Spanish grant FPA2014-52218-P by Xunta de Galicia (GRC2013-024),  by FEDER and by  the Maria de Maeztu Unit of Excellence MDM-2016-0692.  J. M.  P. is supported by the Spanish FPU fellowship FPU14/06300. Centro de F\'{i}sica do Porto is partially funded by FCT through the project CERN/FIS-NUC/0045/2015.

\appendix

\vskip 3cm
\renewcommand{\theequation}{\rm{A}.\arabic{equation}}
\setcounter{equation}{0}

\section{Details of the background}
\label{Background_details}

Let us write a coordinate representation of the internal part of our background. The metric of ${\mathbb C}{\mathbb P}^2$ can be written as:
\beq
ds^2_{{\mathbb C}{\mathbb P}^2}\,=\,
d\chi^2+{\cos^2\chi\over 4}((\omega^1)^2+(\omega^2)^2)\,+\,
{\cos^2\chi \sin^2\chi \over 4} (\omega^3)^2\,\,,
\eeq
where $\chi$ is an angular coordinate taking values in the range $0\le\chi\le \p$ and  $\omega^1$, $\omega^2$ and $\omega^3$ are  three SU(2) left-invariant one-forms, which can be written in terms of three angles $(\theta, \varphi, \psi)$ as follows:
\bear
&&\omega^1\,=\,\cos\psi\,d\theta\,+\,\sin\psi\,\sin\theta\,d\varphi\,\,,\rc\rc
&&\omega^2\,=\,\sin\psi\,d\theta\,-\,\cos\psi\,\sin\theta\,d\varphi\,\,,\rc\rc
&&\omega^3\,=\,d\psi\,+\,\cos\theta\,d\varphi\,\,.
\label{SU2_oneforms}
\eear
The fiber $\tau$ in (\ref{compact_flavored_metric})  takes values in the range $0\le\tau\le 2\pi$ and the one-form $A$  is:
\beq
A\,=\,{1\over 2}\,\cos^2\big({\chi\over 2}\big)\,\omega^3\,\,.
\eeq
The vielbein basis of ${\mathbb C}\,{\mathbb P}^2$ is:
\bear
&&e^1\,=\,{1\over 2}\cos\big({\chi\over 2})\,\omega^1\,\,,
\qquad\qquad\qquad\,\,\,\,\,\,\,
e^2\,=\,{1\over 2}\cos\big({\chi\over 2})\,\omega^2\,\,,\rc\rc
&&e^3\,=\,{1\over 2}\cos\big({\chi\over 2})\,\sin\big({\chi\over 2})\,\omega^3\,\,,
\qquad\qquad
e^4\,=\,{1\over 2}\,d\chi\,\,.
\eear
We can use these one-forms to define  the two-form   $\hat\Omega_2$ as:
\beq
\hat\Omega_2\,=\,e^{3 i\tau}\,(e^1+i e^2)\wedge (e^3+i e^4)\,\,,
\label{hat_Omega_2}
\eeq
Let us now write our ansatz for $F_3$ as:
\beq
F_3\,=\,Q_f\,dx^3\wedge {\rm Im }\,\hat\Omega_2\,\,,
\label{F3_ansatz}
\eeq
where $Q_f$ is  a constant  proportional to the number of flavors $N_f$.
The modified Bianchi identity for $F_3$ is:
\beq
dF_3\,=\,-3 \,Q_f\,dx^3\wedge {\rm Re}\,\hat\Omega_2\wedge (d\tau+A)\,\,.
\label{dF_3}
\eeq
The dilaton for our solution is:
\beq
e^{{3\phi\over 2}}\,=\,{3\over 4\,Q_f}\,r\,\,
\label{dilaton_sol}
\eeq
Moreover, the RR five-form $F_5$ for our background can be written as:
\beq
F_5\,=\,\partial_r\,\big(e^{-\phi}\,h^{-1}\big)\,\big(1+*\big)\,d^4x\wedge dr\,\,.
\label{F5_sol}
\eeq
The precise relation between $ Q_f$ and $ N_f$ can be obtained by analyzing the embeddings of the  family  of flavor branes that source the background. For the case of flavor branes dual to massless quarks we get:
\beq
Q_f\,=\,{4\,\pi\,N_f\over 9\,\sqrt{3}}\,\,.
\label{Qf_Nf}
\eeq

\renewcommand{\theequation}{\rm{B}.\arabic{equation}}
\setcounter{equation}{0}

\section{Wilson loops  and entanglement entropies}
\label{WL-EE}

In this appendix we calculate the potential energy for  static quark-antiquark pairs, as well as the entanglement entropy for slab regions and their complements.

\subsection{Quark-antiquark potentials}
\label{WL}

To calculate the potential energy between a ``quark" and an ``antiquark" we will follow the holographic prescription to compute the Wilson loops  developed in \cite{Maldacena:1998im,Rey:1998ik} .  In this method one has to solve the equations of motion of a fundamental string with  its two ends lying at the UV boundary. These equations are obtained by extremizing the Nambu-Goto action:
\beq
S\,=\,{1\over 2\pi}\,\,\int d\tau d\sigma\,e^{{\phi\over 2}}\,\sqrt{-\det g_2}\,\,,
\label{NG_action_general}
\eeq
where $g_2$ is the Einstein frame induced metric on the worldvolume of the  string. We consider separately the cases in which the quark and the antiquark are in the same layer (\ie\ with the same value of the coordinate $x^3$) and the configuration in which they have the same value of 
$(x^1, x^2)$ and different values of $x^3$. 

\subsubsection{ Intra-layer  potential}
\label{parallel_potential}
Let us first consider a fundamental string hanging from the UV boundary $r\to\infty$ and extended along one of the layer directions (say  along $x^1\equiv x$). with the other two cartesian coordinates being constant.  We parametrize the worldvolume of such a string by means of the coordinates $(\tau, \sigma)=(x^0, x^1)$.  The Nambu-Goto action (\ref{NG_action_general}) takes the form:
\beq
{S\over T}\,=\,\int dx\,e^{{\phi\over 2}}\,\sqrt{(r')^2\,+\,{r^4\over R^4}}\,\equiv \int dx L\,\,,
\eeq
where the prime denotes derivative with respect to $x$,  $T=\int dx^0$ and we have defined an effective lagrangian function $L$. Since $L$ does not depend explicitly on $x$, the Euler-Lagrange equation of motion has the following first integral:
\beq
r'\,{\partial  L\over \partial r'}\,-\,L\,=\,{\rm constant}\,\,,
\eeq
or, more explicitly:
\beq
{r^4\,e^{{\phi\over 2}}\over \sqrt{(r')^2\,+\,{r^4\over R^4}}}\,=\,r_0^2\,R^2\,e^{{\phi_o\over 2}}\,\,,
\label{first_integral_NG_parallel}
\eeq
where $r_0$ is the turning point, \ie\ the minimal value of the coordinate $r$, and $\phi_0=\phi(r=r_0)$. 
It is now straightforward to use (\ref{first_integral_NG_parallel}) to obtain $r'$:
\beq
r'\,=\,\pm\,{r^2\over R^2}\,\sqrt{\Big({r\over r_0}\Big)^4\,e^{\phi-\phi_0}\,-\,1}\,\,,
\eeq
from which we we easily get the  parallel cartesian coordinate $x$ as a function of the holographic coordinate $r$:
\beq
x(r)\,=\,\pm\,{R^2\over r_0}\,\int_1^{{r\over r_0}}\,
{dy\over y^2\,\sqrt{y^{{14\over 3}}\,-\,1}}\,\,.
\eeq
It follows that the quark-antiquark distance $d_{\parallel}$ at the boundary is:
\beq
d_{\parallel}\,=\,{2R^2\over r_0}\,\int_1^{\infty}\,{dy\over y^2\,\sqrt{y^{{14\over 3}}\,-\,1}}\,=\,
{2R^2\sqrt{\pi}\over r_0}\,
{\Gamma\Big({5\over 7}\Big)\over \Gamma\Big({3\over 14}\Big)}\,\,.
\label{d_parallel}
\eeq
Let us now use (\ref{first_integral_NG_parallel}) to compute the on-shell action for this configuration of the fundamental string.  After some calculation we get:
\beq
{S_{on-shell}\over T}\,=\,2\,{1\over 2\pi}\,{e^{{\phi_0\over 2}}\over r_0^2}\,
\int_{r_0}^{r_{\max}}\,
{r^2\,e^{\phi-\phi_0}\,dr\over
\sqrt{\Big({r\over r_0}\Big)^4\,e^{\phi-\phi_0}\,-\,1}}\,\,.
\eeq
Using the value of the dilaton for our background, we obtain:
\beq
{S_{on-shell}\over T}\,=\,{1\over \pi}\,\Big({3\over 4\,Q_f}\Big)^{{1\over 3}}\,r_0^{{4\over 3}}\,
\int_{1}^{{r_{max}\over r_0}}
{y^{{8\over 3}}\,dy\over \sqrt{y^{{14\over 3}}\,-\,1}}\,\,,
\eeq
which is a divergent integral when $r_{\max}\to \infty$. We regularize this  divergence by subtracting the action of two fundamental strings going straight from $r=0$ to the boundary at $r=r_{rmax}$. The resulting finite action dividen by $T$ is identified with the $q\bar q$  potential:
\beq
V_{q\bar q}\,=\,{S_{on-shell}^{reg}\over T}\,=\,{S_{on-shell}\over T}\,-\,{2\over 2\pi}\int_0^{r_{max}}dr\,e^{{\phi\over 2}}\,=\,
{S_{on-shell}\over T}\,-\,{3\over 4\pi}\,\Big({3\over 4\,Q_f}\Big)^{{1\over 3}}\,r_{max}^{{4\over 3}}\,\,.
\eeq
One can easily show that $V_{q\bar q}$ can be rewritten as:
\beq
V_{q\bar q}\,=\,-{1\over \pi}\,\Big({3\over 4\,Q_f}\Big)^{{1\over 3}}\,r_{0}^{{4\over 3}}\,
\Bigg[{3\over 4}\,-\,\int_1^{\infty} dy\,y^{{1\over 3}}\,
\Bigg({y^{{7\over 3}}\over \sqrt{y^{{14\over 3}}-1}}\,-\,1\Bigg)
\Bigg]\,\,.
\eeq
The integral inside the brackets in this last expression can be computed analytically. We get:
\beq
V_{q\bar q}\,=\,-{3\over 4 \sqrt{\pi}}\,\Big({3\over 4\,Q_f}\Big)^{{1\over 3}}\,r_{0}^{{4\over 3}}\,
{\Gamma\Big({5\over 7}\Big)\over \Gamma\Big({3\over 14}\Big)}\,\,.
\eeq
By using the relation (\ref{d_parallel}) we can eliminate $r_0$ in favor of the $q\bar q$ distance $d_{\parallel}$. After some calculation we get:
\beq
V_{q\bar q}\,=\,-\beta_{\parallel} \,\,{Q_c^{{2\over 3}}\over Q_f^{{1\over 3}}}\,\,\,
{1\over d_{\parallel}^{{4\over 3}}}\,\,,
\qquad\qquad
\beta_{\parallel}\,=\,{16\pi^{{1\over 6}}\over 9\,\cdot 5^{{2\over 3}}}
\Bigg({\Gamma\Big({5\over 7}\Big)\over \Gamma\Big({3\over 14}\Big)}\Bigg)^{{7\over 3}}\,\,.
\eeq

\subsubsection{ Inter-layer  potential}
\label{tranaverse_potential}
Let us now repeat the analysis of the previous section for the case in which the fundamentals  are separated  at the boundary in the transverse direction $x^3\equiv z$ to the layers. 
We now take $\tau=x^0$ and $\sigma=z$ and consider an  ansatz  of the form $r=r(z)$. The corresponding Nambu-Goto action becomes:
\beq
{S\over T}\,=\,{1\over 2\pi}\,\int dz\,e^{{\phi\over 2}}\,\sqrt{(r')^2\,+\,e^{-2\phi}\,{r^4\over R^4}}\,\equiv \int dz L\,\,,
\eeq
where now $r'=dr/dz$. Proceeding as in section \ref{parallel_potential}, we get:
\beq
r'\,=\,e^{-\phi}
{r^2\over R^2}\,\sqrt{\Big({r\over r_0}\Big)^4\,e^{\phi_0-\phi}\,-\,1}\,\,,
\eeq
which yields the following function $z=z(r)$:
\beq
z(r)\,=\,\pm\,{R^2\over r_0^{{1\over 3}}}\,
\Big({3\over 4 Q_f}\Big)^{{2\over 3}}\,
\int_1^{{r\over r_0}}\,
{dy\over y^{{4\over 3}}\,\sqrt{y^{{10\over 3}}\,-\,1}}\,\,,
\eeq
as well as the following transverse distance:
\beq
d_{\perp}\,=\,2\,{R^2\over r_0^{{1\over 3}}}\,
\Big({3\over 4 Q_f}\Big)^{{2\over 3}}\,
\int_1^{\infty}\,
{dy\over y^{{4\over 3}}\,\sqrt{y^{{10\over 3}}\,-\,1}}\,=\,
6\,\sqrt{\pi}\,\Big({3\over 4 Q_f}\Big)^{{2\over 3}}\,
{\Gamma\Big({3\over 5}\Big)\over \Gamma\Big({1\over 10}\Big)}\,{R^2\over r_0^{{1\over 3}}}\,\,.
\eeq
The unregulated  on-shell action in this case is given by:
\beq
{S_{on-shell}\over T}\,=\,{e^{{\phi_0\over 2}}\over \pi \,r_0^2}\,
\int_{r_0}^{r_{\max}}\,
{r^2\,dr\over
\sqrt{\Big({r\over r_0^2}\Big)^4\,e^{\phi_0-\phi}\,-\,1}}\,=\,
{1\over \pi}\,\Big({3\over 4 Q_f}\Big)^{{1\over 3}}\,r_0^{{4\over 3}}\,
\int_{1}^{{r_{max}\over r_0}}
{y^2\,dy\over \sqrt{y^{{10\over 3}}\,-\,1}}\,\,,
\eeq
whereas the $q\bar q$ potential is:
\beq
V_{q\bar q}\,=\,-{1\over \pi}\,\Big({3\over 4\,Q_f}\Big)^{{1\over 3}}\,r_{0}^{{4\over 3}}\,
\Bigg[{3\over 4}\,-\,\int_1^{\infty} dy\,y^{{1\over 3}}\,
\Bigg({y^{{5\over 3}}\over \sqrt{y^{{10\over 3}}-1}}\,-\,1\Bigg)
\Bigg]\,\,.
\eeq
By performing the integral in this last expression we arrive at:
\beq
V_{q\bar q}\,=\,-{3\over 4 \sqrt{\pi}}\,\Big({3\over 4\,Q_f}\Big)^{{1\over 3}}\,r_{0}^{{4\over 3}}\,
{\Gamma\Big({3\over 5}\Big)\over \Gamma\Big({1\over 10}\Big)}\,\,.
\eeq
Finally, we can rewrite this  intra-layer potential in terms of $d_{\perp}$ as:

\beq
V_{q\bar q}\,=\,-\beta_{\perp}\,\, {Q_c^{2}\over Q_f^{3}}\,\,\,
{1\over d_{\perp}^{4}}\,\,,
\qquad\qquad
\beta_{\perp}\,=\,{2^{12}\,\pi^{{3\over 2}}\over 3^2\,\cdot 5^{2}}
\Bigg({\Gamma\Big({3\over 5}\Big)\over \Gamma\Big({1\over 10}\Big)}\Bigg)^{5}\,\,.
\eeq

\subsection{Entanglement entropy}

Let $A$ be a spatial region in the gauge theory. The holographic entanglement entropy between $A$ and its complement is obtained by finding the eight-dimensional spatial surface $\Sigma$ whose boundary coincides with the boundary of $A$ and minimizes the functional \cite{Ryu:2006bv,Ryu:2006ef}:
\beq
S_{A}\,=\,{1\over 4 G_{10}}\,\int_{\Sigma} d^8\xi\,\sqrt{\det g_8}\,\,,
\eeq
where  $G_{10}$ is the ten-dimensional Newton constant ($G_{10}=8\pi^6$ in our units) and $g_8$ is the induced metric on $\Sigma$ in the Einstein frame. The entanglement entropy between $A$ and its complement is given by $S_A$ evaluated on the minimal surface $\Sigma$. We will obtain $S_A$ when $A$ is a slab extended infinitely in two spatial  cartesian directions and having a finite width in the third one. We will consider separately the two cases corresponding to the two possible orientations of the slab.

\subsubsection{Parallel slab}

Let us consider first the case in which $A$ is the region 
$\{-{l_{\parallel}\over 2}\,\le\,x^1\,\le {l_{\parallel}\over 2}\,, \,-\infty< x^2, x^3<+\infty\}$, \ie\ when the slab has a finite width in the direction parallel to the layers. We will characterize  the surface $\Sigma$ by a function 
$r=r(x)$, where $x\equiv x^1$. After integrating over all coordinates except $x$, we get:
\beq
{S_{\parallel}\over L_2\,L_3}\,=\,
{R^4\over 32\pi^3}\,\Big({3\over 2\sqrt{2}}\Big)^{6}\,
\int e^{-\phi}\,r\,\sqrt{ (r')^2\,+\,{r^4\over R^4}}\,\,dx\,\,,
\eeq
where $L_{2,3}=\int dx^{2,3}$ and $r'=dr/dx$.  The Euler-Lagrange equations which minimize 
$S_{\parallel}$ admit the following first integral:
\beq
{r^5\,e^{-\phi}\over \sqrt{ (r')^2\,+\,{r^4\over R^4}}}\,=\,R^2\,r_0^3\,e^{-\phi_0}\,\,,
\eeq
where $r_0$ is the minimal value of $r$ and $\phi_0=\phi(r=r_0)$. It follows that $r'$ is given by:
\beq
r'\,=\,\pm\,{r^2\over R^2}\,\sqrt{
\Big({r\over r_0}\Big)^6\,e^{2(\phi_0-\phi)}\,-\,1}\,=\,
\pm\,{r^2\over R^2}\,\sqrt{\Big({r\over r_0}\Big)^{{14\over 3}}\,-\,1}\,\,,
\eeq
and, therefore:
\beq
x(r)\,=\,\pm\,{R^2\over r_0}\,
\int_1^{{r\over r_0}}\,
{dy\over y^2\,\sqrt{y^{{14\over 3}}\,-\,1}}\,\,.
\eeq
Then, the length $l_{\parallel}$ in the direction parallel to the layers is:
\beq
l_{\parallel}\,=\,{2\,R^2\over r_0}\,
\int_1^{\infty}\,
{dy\over y^2\,\sqrt{y^{{14\over 3}}\,-\,1}}\,=\,
{2R^2\sqrt{\pi}\over r_0}\,
{\Gamma\Big({5\over 7}\Big)\over \Gamma\Big({3\over 14}\Big)}\,\,.
\label{l_parallel}
\eeq
One can now evaluate the entropy for this configuration. We get:
\beq
{S_{\parallel}\over L_2\,L_3}\,=\,{R^4\over 16\,\pi^3}\,\Big({3\over 2\sqrt{2}}\Big)^{6}\,
r_0^2\,e^{-\phi_0}\,\int_1^{{r_{max}\over r_0}}\,
{y^{{8\over 3}}\,dy\over \sqrt{y^{{14\over 3}}\,-\,1}}\,\,.
\label{S_parallel_total}
\eeq
The integral (\ref{S_parallel_total}) is divergent at the UV and has been regulated  by introducing a maximal radial coordinate $r_{\max}$. The divergent part of $S_{\parallel}$ can be obtained by computing the contribution of the upper limit to the integral  (\ref{S_parallel_total})  and gives:
\beq
{S_{\parallel}^{div}\over L_2\,L_3}\,=\,
{3\,R^4\over 64\,\pi^3}\,\Big({3\over 2\sqrt{2}}\Big)^{6}\,
\Big({4\,Q_f\over 3}\Big)^{{2\over 3}}\,r_{max}^{{4\over 3}}\,\,.
\eeq
We now define $S_{\parallel}^{finite}$ as:
\beq
{S_{\parallel}^{finite}\over L_2\,L_3}\,=\,
{S_{\parallel}\,-\,S_{\parallel}^{div}\over L_2\,L_3}\,\,.
\eeq
One can readily demonstrate that:
\beq
{S_{\parallel}^{finite}\over L_2\,L_3}\,=\,-
{R^4\over 16\,\pi^3}\,\Big({3\over 2\sqrt{2}}\Big)^{6}\,
r_0^2\,e^{-\phi_0}\,
\Bigg[{3\over 4}\,-\,\int_1^{\infty} dy\,y^{{1\over 3}}\,
\Bigg({y^{{7\over 3}}\over \sqrt{y^{{14\over 3}}-1}}\,-\,1\Bigg)
\Bigg]\,\,,
\eeq
which, after performing the integration, gives:
\beq
{S_{\parallel}^{finite}\over L_2\,L_3}\,=\,-{3\sqrt{\pi}\over 64\pi^3}\,
\Big({3\over 2\sqrt{2}}\Big)^{6}\,
\Big({4\,Q_f\over 3}\Big)^{{2\over 3}}
{\Gamma\Big({5\over 7}\Big)\over \Gamma\Big({3\over 14}\Big)}
R^4\,r_0^{{4\over 3}}\,\,.
\eeq
By using the relation  (\ref{l_parallel}) between $r_0$ and $l_{\parallel}$, we can rewrite 
 $S_{\parallel}^{finite}$  as:
\beq
{S_{\parallel}^{finite}\over L_2\,L_3}\,=\,-\gamma_{\parallel}\,
{Q_f^{{2\over 3}}\,Q_c^{{5\over 3}}\over l_{\parallel}^{{4\over 3}}}\,\,,
\qquad\qquad
\gamma_{\parallel}\,=\,{2\over 45\cdot 5^{{2\over 3}}\,\pi^{{11\over 6}}}
\Bigg({\Gamma\Big({5\over 7}\Big)\over \Gamma\Big({3\over 14}\Big)}\Bigg)^{{7\over 3}}\,\,.
\eeq

\subsubsection{Transverse slab}

We now take $A$ to be 
$\{-\infty< x^1, x^2<+\infty\,,\,-{l_{\perp}\over 2}\,\le\,x^3\,\le {l_{\perp}\over 2}\}$, \ie\ a slab with finite width in the direction $x^3$ transverse to the layers. If $z\equiv x^3$, the surface $\Sigma$ is parametrized by a function $r=r(z)$ and the functional to be minimized is:
\beq
{S_{\perp}\over L_1\,L_2}\,=\,
{R^4\over 32\pi^3}\,\Big({3\over 2\sqrt{2}}\Big)^{6}\,\int r\,\sqrt{ (r')^2\,+\,{r^4\over R^4}\,e^{-2\phi}\,}\,\,dz\,\,.
\eeq
The corresponding first integral is now:
\beq
{r^5\,e^{-2\phi}\over \sqrt{ (r')^2\,+\,{r^4\over R^4}\,e^{-2\phi}}}\,=\,R^2\,r_0^3\,e^{-\phi_0}\,\,,
\eeq
and, as a consequence, $r'$ is given by:
\beq
r'\,=\,\pm\,{r^2\over R^2}\,e^{-\phi}\,
\sqrt{
\Big({r\over r_0}\Big)^6\,e^{2(\phi_0-\phi)}\,-\,1}\,=\,
\pm\,{r^2\over R^2}\,e^{-\phi}\,\sqrt{\Big({r\over r_0}\Big)^{{14\over 3}}\,-\,1}\,\,.
\eeq
Therefore $z(r)$ is the following integral:
\beq
z(r)\,=\,\pm\,\Big({3\over 4 Q_f}\Big)^{{2\over 3}}\,
{R^2\over r_0^{{1\over 3}}}\,
\int_1^{{r\over r_0}}\,
{dy\over y^{{4\over 3}}\,\sqrt{y^{{14\over 3}}\,-\,1}}\,\,,
\eeq
and the transverse length $l_{\perp}$ is related to $r_0$ as:
\beq
l_{\perp}\,=\,6\,\sqrt{\pi}\,\Big({3\over 4 Q_f}\Big)^{{2\over 3}}\,
{\Gamma\Big({4\over 7}\Big)\over \Gamma\Big({1\over 14}\Big)}\,{R^2\over r_0^{{1\over 3}}}\,\,.
\label{l_perp}
\eeq
The functional $S_{\perp}$ evaluated on the minimal surface is given by:
\beq
{S_{\perp}\over L_1\,L_2}\,=\,
{R^4\over 16\,\pi^3}\,\Big({3\over 2\sqrt{2}}\Big)^{6}\,r_0^{2}\,
\int_1^{{r_{max}\over r_0}}\,
{y^{{10\over 3}}\,dy\over \sqrt{y^{{14\over 3}}\,-\,1}}\,\,,
\eeq
and its divergent part is:
\beq
{S_{\perp}^{div}\over L_1\,L_2}\,=\,{R^4\over 32\,\pi^3}\,\Big({3\over 2\sqrt{2}}\Big)^{6}\,r_{max}^{2}\,\,.
\eeq
Defining $S_{\perp}^{finite}$ by subtracting $S_{\perp}^{div}$ from $S_{\perp}$:
\beq
{S_{\perp}^{finite}\over L_1\,L_2}\,=\,
{S_{\perp}\,-\,S_{\perp}^{div}\over L_1\,L_2}\,\,,
\eeq
we get:
\beq
{S_{\perp}^{finite}\over L_1\,L_2}\,=\,-
{R^4\over 16\,\pi^3}\,\Big({3\over 2\sqrt{2}}\Big)^{6}\,r_{0}^{2}
\Bigg[{1\over 2}\,-\,
\int_1^{\infty} dy\,y\,
\Bigg({y^{{7\over 3}}\over \sqrt{y^{{14\over 3}}-1}}\,-\,1\Bigg)
\Bigg]\,\,,
\eeq
which, after computing the integral, becomes: 
\beq
{S_{\perp}^{finite}\over L_1\,L_2}\,=\,-
{1\over 32\,\pi^3}\,\Big({3\over 2\sqrt{2}}\Big)^{6}\,\sqrt{\pi}\,
{\Gamma\Big({4\over 7}\Big)\over \Gamma\Big({1\over 14}\Big)}\,
R^4\,r_{0}^{2}\,\,.
\eeq
Finally, using the relation (\ref{l_perp}), we arrive at:
\beq
{S_{\perp}^{finite}\over L_1\,L_2}\,=\,-\gamma_{\perp}\,{Q_c^4\over Q_f^4}\,\,
{1\over l_{\perp}^6}\,\,,
\qquad\qquad
\gamma_{\perp}\,=\,\Big({16\over 15}\Big)^4\,\sqrt{\pi}
\Bigg({\Gamma\Big({4\over 7}\Big)\over \Gamma\Big({1\over 14}\Big)}\Bigg)^{7}\,\,.
\eeq

\renewcommand{\theequation}{\rm{C}.\arabic{equation}}
\setcounter{equation}{0}

\section{More on the reduced equations}
\label{Reduced_details}
In this appendix we give details on the dimensional reduction of our setup. We first consider the reduction to four dimensions.

\subsection{4d reduction}

Let us consider the   reduction ansatz of the 10d metric written in (\ref{4d_metric_ansatz}).
For this ansatz, the determinant of the 10d and 4d metrics are related as:
\beq
\sqrt{-G_{10}}\,=\,e^{{10\over 3}\gamma-\beta}\,\sqrt{G_5}\,\sqrt{-g_4}\,\,,
\eeq
where $G_5$ is the determinant of the 5d compact internal manifold. Moreover,  the relation between the Ricci scalars in 10d and 4d is:
\beq
R_{10}\,=\,e^{-{10\over 3}\gamma+\beta}\,
\Big[R_4\,-\,{40\over 3}\,(\partial\gamma)^2\,-\,20\,(\partial\lambda)^2\,-\,{3\over 2}\,(\partial\beta)^2\,+\,
24\,e^{{16\over 3}\gamma+2\lambda-\beta}\,-\,4\,e^{{16\over 3}\gamma+12\lambda-\beta}\,+\,
\Lambda\Big]\,\,,
\eeq
where $\Lambda$ is given by:
\beq
\Lambda\,=\,{1\over \sqrt{-g_4}}\partial_m\,
\Big[\,\sqrt{-g_4}\,g^{mn}\,\partial_n\,\Big(\beta-{10\over 3}\gamma\Big)\Big]\,\,.
\eeq
As $\Lambda$  leads to a total derivative in the 4d  Einstein-Hilbert action and, thus, it 
 does not contribute to the equations of motion and we simply drop it from our equations. 
The Einstein-Hilbert action in 10d can be written as:
\bear
&&\int d^{10}X\, \sqrt{-G_{10}}\,R_{10}\,=\,V_5\,V_{x^3}\,
\int d^4z \sqrt{-g_4}\,\Big[
R_4\,-\,{40\over 3}\,(\partial\gamma)^2\,-\,20\,(\partial\lambda)^2\,-\,{3\over 2}\,(\partial\beta)^2\,+\,
\rc\rc
&&\qquad\qquad\qquad\qquad\qquad\qquad\qquad\qquad
+\,24\,e^{{16\over 3}\gamma+2\lambda-\beta}\,-\,4\,e^{{16\over 3}\gamma+12\lambda-\beta}\Big]\,\,,
\label{EH-action}
\eear
where $V_5$ is the volume of the five-dimensional compact space and $V_{x^3}\equiv \int dx^3$.

Let us now write the contribution of the remaining fields of type IIB supergravity to the reduced action. We start with the contribution of the dilaton $\phi$, which is proportional to:
\beq
\int d^{10} X\,\sqrt{-G_{10}}\,\,{1\over 2}\,G^{MN}\,\partial_M\,\phi\,\partial_N\,\phi\,=\,
V_5\,V_{x^3}\,\int d^4z\,\sqrt{-g_4}\,\,\,{1\over 2}\,
g^{mn}\,\partial_m\,\phi\,\partial_n\phi\,\,.
\eeq
Moreover, the RR five-form $F_5$ in these new variables is:
\beq
F_5\,=\,Q_c\,e^{{40\over 3}\gamma\,-\,\beta}\,\sqrt{-g_4}\,\,d^4z\,\wedge\,dx^3\,\,,
\eeq
and its contribution to the effective action is proportional to:
\beq
\int {1\over 2}\,F_5\wedge {}^* F_5\,=\,V_5\,V_{x^3}\,\int d^4z \sqrt{-g_4}\,\,\,
{Q_c^2 \over 2}\,e^{{40\over 3}\gamma\,-\,\beta}\,\,.
\eeq
Similarly, the RR three-form $F_3$ contributes as:
\beq
\int {1\over 2}\,e^{\phi}\,F_3\wedge {}^* F_3\,=\,V_5\,V_{x^3}\,\int d^4z \sqrt{-g_4}\,\,\,
Q_f^2 \,e^{\phi+4\gamma+4\lambda-3\beta}\,\,.
\eeq
It remains to calculate  the contribution of the DBI action of the flavor D5-branes, which is given by:
\beq
-{3 V_5\,V_{x^3}\,Q_f\over \,\kappa_{10}^2}\,\int d^4z\,
\sqrt{-g_4}\,\,e^{{14\over 3}\gamma\,-\,2\beta\,-2\lambda\,+\,{\phi\over 2}}\,\,.
\eeq
Putting everything together, we can write the effective action as in (\ref{effective_4d_action}), 
where $V$ is the  potential for the scalar fields $\phi$, $\gamma$, $\lambda$ and $\beta$ written in 
(\ref{4d_potential}). 

Let us now write down the equations of motion derived from the action (\ref{effective_4d_action}). First of all, the equation of motion for the 4d metric is:
\beq
R_{mn}\,=\,{1\over 2}\,\partial_m\phi\,\partial_n \phi\,+\,
{40\over 3}\,\partial_m\gamma\,\partial_n \gamma\,+\,
20\, \partial_m\lambda\,\partial_n \lambda\,+\,
{3\over 2}\partial_m\beta\,\partial_n \beta\,+\,{1\over 2}\,g_{mn}\,V\,\,,
 \label{eoms_4d_metric}
\eeq
where $R_{mn}$ is the Ricci tensor for $g_{mn}$.  These equations are equivalent to the ones written 
in (\ref{Einstein_eq_Psi}). Moreover, if we define the d'Alembertian of any scalar field $\Psi$ as in (\ref{4d_Dalembertian}), the equations for $\phi$, $\gamma$, $\lambda$ and $\beta$ are:
\bear
&& \Box \,\phi\,=\,\partial_{\phi} V\,\,,
\qquad\qquad\qquad
 \Box \,\gamma\,=\,{3\over 80}\,\,\partial_{\gamma} V\,\,,\rc\rc
 &&\Box \,\lambda\,=\,{1\over 40}\,\partial_{\lambda} V\,\,,
 \qquad\qquad\qquad
 \Box \,\beta\,=\,{1\over 3}\,\partial_{\beta} V\,\,,
 \label{eoms_4d_scalar}
 \eear
where we have denoted ${\partial  V\over \partial \phi}=\partial_{\phi}\,V$ and similarly for the other scalar fields. Notice that the four equations in (\ref{eoms_4d_scalar}) can be written  more compactly as in
(\ref{scalar_eom_4d_compact}). Let us now  write the equations (\ref{eoms_4d_scalar}) for the scalars more explicitly:
\bear
&& \Box \,\phi\,=\,Q_f^2\,e^{4\gamma+4\lambda-3\beta+\phi} \,+\,3 Q_f\,e^{{14\over 3}\gamma - 2\lambda - 2\beta\,+\,{\phi\over 2}}\,\,,\rc\rc
&&\Box \,\gamma\,=\,-{24\over 5}\,e^{{16\over 3}\gamma+2\lambda-\beta}\,+\,
{4\over 5}\,e^{{16\over 3}\,\gamma+12 \lambda-\beta}\,+\,{3 \,Q_f^2\over 20}\,
e^{4\gamma+4\lambda-3\beta+\phi}\,+\,{Q_c^2\over 4}\,
e^{{40\over 3}\gamma-\beta} \,+\,
{21 Q_f\over 20}\,e^{{14\over 3}\gamma - 2\lambda - 2\beta\,+\,{\phi\over 2}}\,\,,\rc\rc
&& \Box \,\lambda\,=\,-{6\over 5}\,\,e^{{16\over 3}\gamma+2\lambda-\beta}\,+\,{6\over 5}\,
e^{{16\over 3}\,\gamma+12 \lambda-\beta}\,+\,{Q_f^2\over 10}\,e^{4\gamma+4\lambda-3\beta+\phi} \,-\,
{3\,Q_f\over 10}\,e^{{14\over 3}\gamma - 2\lambda - 2\beta\,+\,{\phi\over 2}}\,\,,\rc\rc
&& \Box \,\beta\,=\,8\,e^{{16\over 3}\gamma+2\lambda-\beta}-{4\over 3}
e^{{16\over 3}\,\gamma+12 \lambda-\beta}-Q_f^2
e^{4\gamma+4\lambda-3\beta+\phi} -{Q_c^2\over 6}
e^{{40\over 3}\gamma-\beta}-4\,Q_f e^{{14\over 3}\gamma - 2\lambda - 2\beta\,+\,{\phi\over 2}}\,\,.
\qquad
\label{eom_scalars_4d_explicit}
\eear

\subsection{5d reduction}

Let us now consider a reduction of the 10d metric to a 5d metric according to the ansatz (\ref{10d_5d_metric_ansatz}). 
The determinants of the 10d and 5d metrics are related as:
\beq
\sqrt{-G_{10}}\,=\,e^{{10\over 3}\gamma}\,\sqrt{G_5}\,\sqrt{-g_5}\,\,,
\eeq
where $G_5$ is the determinant of the 5d compact internal manifold.  Up to terms which give a total derivative in the Einstein-Hilbert action, the Ricci scalars in 10d and 5d are related as:
\beq
R_{10}\,=\,e^{-{10\over 3}\gamma}\,
\Big[R_5\,-\,{40\over 3}\,(\partial\gamma)^2\,-\,20\,(\partial\lambda)^2\,+\,
24\,e^{{16\over 3}\gamma+2\lambda}\,-\,4\,e^{{16\over 3}\gamma+12\lambda}\,\Big]\,\,.
\eeq
Then, the Einstein-Hilbert action in 10d can be written as:
\beq
\int d^{10}X\, \sqrt{-G_{10}}\,R_{10}=V_5
\int d^5z \sqrt{-g_5}\,\Big[
R_5\,-\,{40\over 3}\,(\partial\gamma)^2\,-\,20\,(\partial\lambda)^2
+24\,e^{{16\over 3}\gamma+2\lambda}-4\,e^{{16\over 3}\gamma+12\lambda}\Big]\,\,,
\label{EH-action_5d}
\eeq
where $V_5$ is the volume of the 5d compact space.   Let us write the contribution of the remaining fields of type IIB supergravity  to the effective action. The dilaton contributes as:
\beq
\int d^{10} X\,\sqrt{-G_{10}}\,\,{1\over 2}\,G^{MN}\,\partial_M\,\phi\,\partial_N\,\phi\,=\,
V_5\,\int d^5z\,\sqrt{-g_5}\,\,\,{1\over 2}\,\,
g^{pq}\,\partial_p\,\phi\,\partial_q\phi\,\,.
\eeq
The RR five-form is:
\beq
F_5\,=\,Q_c\,e^{{40\over 3}\gamma}\,\sqrt{-g_5}\,\,d^5z\,,
\eeq
and contributes to the effective action as:
\beq
\int {1\over 2}\,F_5\wedge {}^* F_5\,=\,V_5\,\int d^5z \sqrt{-g_5}\,\,\,
{Q_c^2 \over 2}\,e^{{40\over 3}\gamma}\,\,.
\eeq
Let us consider the following ansatz for the RR three-form $F_3$:
\beq
F_3\,=\,{1\over \sqrt{2}}\,{\cal F}_1\wedge {\rm Im}\, \hat\Omega_2\,\,,
\eeq
where $\hat\Omega_2$ is the two-form (\ref{hat_Omega_2}) and ${\cal F}_1$ has only components along the 5d space. We will represent ${\cal F}_1$  in terms of a scalar potential ${\cal V}$ as in (\ref{F1-V}). Then, the contribution of ${\cal V}$ to the action is:
\beq
{1\over 2}\,\int_{{\cal M}_{10}}\,F_3\wedge {}^*F_3\,=\,
{V_5\over 2}\,\int\,d^5 z\,\sqrt{-g_5}\,e^{4\gamma+4\lambda+\phi}\,(\partial {\cal V})^2\,\,.
\eeq
The DBI action of the flavor D5-branes is:
\beq
S_{DBI}\,=\,-T_5\,\sum_{N_f}\,\int d^6\xi\,e^{{\phi\over 2}}\,\sqrt{-\hat g_6}
\eeq
After smearing and integration over the internal manifold, the DBI action becomes:
\beq
S_{DBI}\,=\,-{6\,Q_f\,V_5\over 2\kappa_{10}^2}\,
\int d^5 z\,\sqrt{-\hat g_4}\,e^{{\phi\over 2}+{14\over 3}\,\gamma-2\lambda}\,\,,
\eeq
where $\hat g_4$ is the determinant of the metric obtained by taking the pullback of the 5d metric on a surface  with constant  $x^3$. Putting everything together we arrive at the effective action (\ref{effective_5d_action}). The equations of motion for the scalars
$\phi$, $\gamma$ and $\lambda$  derived from (\ref{effective_5d_action}) are:
\bear
&&\Box \,\phi\,=\,\partial_{\phi} U\,+\,{1\over 2}\,e^{4\lambda+4\gamma+\phi}\,
\big(\partial{\cal V}\big)^2\,+\,3\,Q_f\,{\sqrt{-\hat g_4}\over \sqrt{-g_5}}\,
e^{{14\over 3}\gamma\,-\,2\lambda\,+\,{\phi\over 2}}\,\,,\rc\rc
&&\Box \,\gamma\,=\,{3\over 80}\,\partial_{\gamma} U
\,+\,{3\over 40}\,e^{4\lambda+4\gamma+\phi}\,
\big(\partial{\cal V}\big)^2\,+\,{21\over 20}\,Q_f\,{\sqrt{-\hat g_4}\over \sqrt{-g_5}}\,
e^{{14\over 3}\gamma\,-\,2\lambda\,+\,{\phi\over 2}}\,\,,\rc\rc
&&\Box \,\lambda\,=\,{1\over 40}\,\partial_{\lambda} U
\,+\,{1\over 20}\,e^{4\lambda+4\gamma+\phi}\,
\big(\partial{\cal V}\big)^2\,
-\,{3\over 10}\,Q_f\,{\sqrt{-\hat g_4}\over \sqrt{-g_5}}\,
e^{{14\over 3}\gamma\,-\,2\lambda\,+\,{\phi\over 2}}\,\,,
\label{5d_scalar_eoms}
\eear
where $\Box$ is the laplacian operator for the 5d metric.  Let us group the 5d scalars into a single 
three-component field  $\Psi=(\phi, \gamma, \lambda)$. Then, the three scalar equations of 
(\ref{5d_scalar_eoms}) can be compactly written as:
\beq
\Box \Psi\,=\,\alpha_{\Psi}\,\partial_{\Psi}\,U\,+\,{1\over 2}\,\alpha_{\Psi}\,\big(\partial {\cal V}\big)^2\,\partial_{\Psi}\, \Big(e^{4\lambda+4\gamma+\phi}\Big)
\,+\,6\,Q_f\,
{\sqrt{-\hat g_4}\over \sqrt{-g_5}}\,\alpha_{\Psi}\,\partial_{\Psi}
\Big( e^{{14\over 3}\gamma-2\lambda+{\phi\over 2}}\Big)\,\,,
\label{5d_scalar_eoms_compact}
\eeq
where the coefficients $\alpha_{\Psi}$ are those written in (\ref{alpha_Psi}) for the three scalars $(\phi, \gamma, \lambda)$. The equation of ${\cal V}$ is:
\beq
\partial_p\Big[\,\sqrt{-g_5}\,e^{4\lambda+4\gamma+\phi}\, g^{pq}\,\partial_q\,
{\cal V}\,\Big]\,=\,0\,\,,
\label{V_eom}
\eeq
while the Einstein equations are:
\bear
&&R_{pq}\,=\,\sum_{\Psi}\,{1\over 2\alpha_{\Psi}}\,
\partial_p\,\Psi\,\partial_q\,\Psi\,+{1\over 2}\,e^{4\lambda+4\gamma+\phi}\,
\partial_{p}{\cal V}\,\partial_{q}{\cal V}\,+\,{1\over 3}\,g_{pq}\,U\,+\,\rc\rc
&&\qquad\qquad\qquad\qquad
+\,3Q_f\,{\sqrt{-g_4}\over \sqrt{-g_5}}\,e^{{14\over 3}\gamma\,-\,2\lambda\,+\,{\phi\over 2}}\,
\Big({4\over 3}\,g_{pq}\,-\,\hat g^{(4)}_{pq}\Big)\,\,.
\label{Einstein_eom_5d}
\eear
It is straightforward to demonstrate that the metric written in (\ref{5d_metric_ansatz}) and (\ref{d_5dmetric}), together with the scalars displayed in  (\ref{5d_scalars_solution}) and (\ref{cal_V_5d}), satisfy (\ref{5d_scalar_eoms}), (\ref{V_eom}) and (\ref{Einstein_eom_5d}).

\subsection{5d $\to$ 4d reduction}
\label{5d_to_4d}

Let us now perform an additional reduction of the 5d action to four dimensions. We will reduce along the coordinate $x^3$ and we will adopt the following ansatz for the 5d metric:
\beq
ds_5^2\,=\,e^{-\beta}\,ds_4^2\,+\,e^{2\beta}\,(dx^3)^2\,\,,
\eeq
where $\beta$ is a new scalar which depends on the 4d coordinates. The determinant of the 5d metric and the one corresponding to the pullback to the surface $x^3={\rm constant}$ are related to the determinant $g_4$ of the  reduced 4d metric as:
\beq
\sqrt{-g_5}\,=\,e^{-\beta}\,\sqrt{-g_4}\,\,,\qquad\qquad
\sqrt{-\hat g_4}\,=\,e^{-2\beta}\,\sqrt{-g_4}\,\,.
\eeq
Moreover, after neglecting a total derivative, we can relate the Einstein-Hilbert term of the action 
(\ref{EH-action_5d})  to the one corresponding to the reduced 4d action as:
\beq
\int d^5z\,\sqrt{-g_5}\,R_5\,=\,\int d^4z\,\sqrt{-g_4}\,
\big(R_4\,-\,{3\over 2}\,(\partial \beta)^2\big)\,\,.
\eeq
Let us now split the one-form ${\cal F}_1$ as:
\beq
{\cal F}_1\,=\,\chi\,dx^3\,+\,f_1\,\,,
\eeq
where $f_1$ is a closed one-form that has legs only in the 4d space. Using that:
\beq
{\cal F}_1^2\,=\,\big(\partial {\cal V}\big)^2\,=\,e^{-2\beta}\,\chi^2\,+\,e^{\beta}\,f_1^2\,\,,
\eeq
we can write the term containing ${\cal V}$ in (\ref{effective_5d_action}) as:
\beq
 \sqrt{-g_5}\,\Big[-
{1\over 2}\,e^{4\gamma+4\lambda+\phi}\,(\partial {\cal V})^2\Big]\,=\,
 \sqrt{-g_4}\,\Big[-{1\over 2}\,e^{4\gamma+4\lambda+\phi}\,
 \Big(e^{-3\beta}\,\chi^2\,+\,f_1^2\Big)\Big]\,\,.
\eeq
Collecting all these results, we can write the effective action as:
\bear
&&S_{eff}\,=\,{V_5\,V_{x^3}\over 2\,\kappa_{10}^2}\,
\int d^4z\, \sqrt{-g_4}\,\,\Big[
R_4\,-\,{40\over 3}\,(\partial\gamma)^2\,-\,20\,(\partial\lambda)^2\,-\,{3\over 2}\,(\partial\beta)^2\,-\,\rc\rc
&&\qquad\qquad
-{1\over 2}\,(\partial\phi)^2-{1\over 2}\,
e^{4\gamma+4\lambda+\phi}\,f_1^2\,-\,V\,\Big]\,\,,
\label{4d_action_from_5d}
\eear
where the 4d potential $V$ is related to the 5d one $U$ in (\ref{5d_potential}) by the relation:
\beq
V\,=\,e^{-\beta}\,U\,+\,{1\over 2}\,
e^{4\gamma+4\lambda+\phi-3\beta}\,\chi^2\,+\,6\,Q_f\,e^{{14\over 3}\,\gamma-2\lambda-2\beta+{\phi\over 2}}\,\,.
\eeq
It is now straightforward to verify that the action (\ref{4d_action_from_5d}) reduces to the one written in (\ref{effective_4d_action}) when we truncate the former in such a way that $f_1=0$ and  the scalar $\chi$ takes the following constant value:
\beq
\chi\,=\,\sqrt{2}\,Q_f\,\,.
\label{chi_F_1}
\eeq

\renewcommand{\theequation}{\rm{D}.\arabic{equation}}
\setcounter{equation}{0}

\section{Hydrodynamic fluctuations}
\label{hydro_details}

In this appendix we provide details of the analysis of the hydrodynamic fluctuations, which complement the presentation given in section \ref{hydro} on the main text. We will  consider separately the two channels.

\subsection{Shear channel}
One can show that the fluctuation equations (\ref{fluct_eq_Psi}) and (\ref{fluct_eq_h}) for the ansatz (\ref{shear_ansatz}) reduce to:
\bear
&&H_{tx}''\,+\,\partial_r\log\Big({c_2^4\over c_1\,c_3}\Big)\,H_{tx}'\,+\,
W\,H_{tx}\,-\,\,q\,{c_3^2\over c_2^2}\,\big(q\,H_{tx}\,+\,\omega\,H_{xy}\big)\,=\,0\,\,,\rc\rc
&&H_{xy}''\,+\,\partial_r\log\Big({c_1\,c_2^2\over c_3}\Big)\,H_{xy}'\,+\,
 W\,H_{xy}\,+\,\,\omega\,{c_3^2\over c_1^2}\,\big(q\,H_{tx}\,+\,\omega\,H_{xy}\big)\,=\,0\,\,,\rc\rc
&&q\,c_1^2\,H_{xy}'\,+\,\omega\,c_2^2\,H_{tx}'\,=\,0\,\,,
\label{shear_fluct}
\eear
where $ W$ is the function:
\beq
 W\,=\,c_3^2\,\,V\,+\,2\,\partial_r^2\log c_2\,+\,
 2\,\partial_r\log c_2\,\partial_r
\log\Big({c_1\,c_2^2\over c_3}\Big)\,\,.
\label{W_def}
\eeq
One can verify easily that  ${ W}$  vanishes for our background. Therefore, we will omit it in the equations that follow in this section. Notice that the last equation in (\ref{shear_fluct}) is first-order in the radial derivative and it  can be used to reduce the number of equations of the system.  Actually, if we 
define the gauge invariant combination $X$ as in (\ref{X_def})
and combine this definition and the last equation in (\ref{shear_fluct}) to express 
the first derivatives of $H_{tx}$ and $H_{xy}$ in terms of $X'$, we  get:
\beq
H_{tx}'\,=\,{q\,c_1^2\over q^2\,c_1^2\,-\,\omega^2\,c_2^2}\,\,X'\,\,,
\qquad\qquad
H_{xy}'\,=\,-{\omega\,c_2^2\over q^2\,c_1^2\,-\,\omega^2\,c_2^2}\,\,X'\,\,.
\label{H_X}
\eeq
Moreover, one can show that the system (\ref{shear_fluct}) reduces to the following second-order differential equation for $X$:
\bear
X''\,+\,{q^2\,c_1^2\,\partial_r\log\Big({c_2^4\over c_1\,c_3}\Big)\,-\,
\omega^2\,c_2^2\,\partial_r\log\Big({c_1\,c_2^2\over c_3}\Big)\over
q^2\,c_1^2\,-\,\omega^2\,c_2^2}\,X'\,-\,
{c_3^2\over c_2^2\,c_1^2}\,(q^2\,c_1^2\,-\,\omega^2\,c_2^2)\,X\,=\,0\,\,. 
\qquad\qquad
\label{eom_X_cs}
\eear
By using the values of $c_1$, $c_2$ and $c_3$ written in (\ref{cs_4d_metric}), one can easily demonstrate that  (\ref{eom_X_cs})  can  be converted into (\ref{eom_X}). 

\subsection{Sound channel}
Plugging (\ref{scalar_wawes_compact}) and (\ref{metric_waves}) into (\ref{fluct_eq_Psi}) we get the following second order equation for $\hat\Psi(r)$:
\bear
&&\hat\Psi''\,+\,\partial_r\log\Big({c_1\,c_2^2\over c_3}\Big)\,\hat\Psi'\,+\,
\Big[{c_3^2\over c_1^2}\,\omega^2\,-\,{c_3^2\over c_2^2}\,q^2\Big]\,\hat\Psi\,+\,
\rc\rc
&&\qquad\qquad\qquad\qquad
+{1\over 2}\Psi'\,(H'_{xx}+H'_{yy}-H'_{tt})\,=\,c_3^2\,\alpha_{\Psi}\,
\hat\delta\big[\partial_{\Psi}\,V\big]
\eear
where, for every $\Psi=(\phi, \gamma, \lambda, \beta)$, we define:
\beq
\hat\delta\big[\partial_{\Psi}\,V\big]\,=\,\partial_{\phi}\partial_{\Psi}V\,\Phi(r)\,+\,
\partial_{\gamma}\partial_{\Psi}V\,\Gamma(r)\,+\,
\partial_{\lambda}\partial_{\Psi}V\,\Lambda(r)\,+\,
\partial_{\beta}\partial_{\Psi}V\,B(r)\,\,.
\eeq
Let us now write the equations for the metric fluctuations, which are obtained by taking different values for the $(m,n)$ indices in (\ref{fluct_eq_h}). To write these equations compactly, let us denote by $\hat\delta V$  the following radial function:
\beq
\hat\delta V\,=\,\partial_{\phi}V\,\Phi(r)\,+\,
\partial_{\gamma}V\,\Gamma(r)\,+\,
\partial_{\lambda}V\,\Lambda(r)\,+\,
\partial_{\beta}V\,B(r)\,\,.
\eeq
Then, one can check that  (\ref{fluct_eq_h}) is equivalent to the following 
second-order equations:
\bear
&&H_{tt}''\,+\,\partial_r\log\Big({c_1^2c_2^2\over c_3}\Big)\,H'_{tt}\,-\,\omega^2\,{c^2_3\over c_1^2}\,
(H_{xx}+H_{yy})-\,q^2\,{c^2_3\over c_2^2}H_{tt}-2\omega\,q\,{c^2_3\over c_1^2}H_{ty}\,-\,\qquad\qquad\rc\rc
&&\qquad\qquad
-\partial_r\,\log c_1\,(H_{xx}'+H_{yy}')\,-\,c_3^2\,\hat\delta V+H_{tt}\,\tilde W=0\,\,,\rc\rc
&&H_{ty}''\,+\,\partial_r\log\Big({c_2^4\over c_1\,c_3}\Big)\,H'_{ty}\,+\,
\omega\,q\,{c_3^2\over c_2^2}\,H_{xx}\,+\,H_{ty}\,W\,=\,0,\,,\rc\rc
&&H_{xx}''+\partial_r\log\Big({c_1 c_2^3\over c_3}\Big)\,H'_{xx}+
\Big(\omega^2{c_3^2\over c_1^2}-q^2\,{c_3^2\over c_2^2}\Big)\,H_{xx}-
\partial_r \log c_2(H_{tt}'-H_{yy}')+c_3^2\,\hat\delta V+H_{xx}\,W=0\,\,,\rc\rc
&&H_{yy}''+\partial_r\log\Big({c_1 c_2^3\over c_3}\Big)\,H'_{yy}+
{c_3^2\over c_1^2}\,(\omega^2H_{yy}+2q\omega H_{ty})\,+\,
q^2\,{c_3^2\over c_2^2}\,(H_{tt}-H_{xx})\,+\,\rc\rc
&&\qquad\qquad\qquad
+\partial_r\,\log c_2\,(H_{xx}'-H_{tt}')\,+\,c_3^2\,\hat\delta V
+H_{yy}\,W=0\,\,,
\label{second_order_metric_fluct}
\eear
together with three first order constraints associated to the gauge fixing condition (\ref{radial_gauge}):
\bear
&&q\,H_{ty}'\,+\,\omega\,(H_{xx}'+H_{yy}')\,=\,\partial_r\log{c_1\over c_2}\,\big(2q\,H_{ty}\,+\,
\omega (H_{xx}+H_{yy})\big)\,-\,\omega\,\sum_{\Psi}\,{\Psi'\over \alpha_{\Psi}}\,\hat \Psi\,\,,\rc\rc
&&\omega\,{c_2^2\over c_1^2}\,H_{ty}'\,+\,q(H_{tt}'-H_{xx}')\,=\,-
q\,\partial_r\log{c_1\over c_2}\,H_{tt}\,+\,q\,\sum_{\Psi}\,{\Psi'\over \alpha_{\Psi}}\,\hat \Psi\,\,,\rc\rc
&&\partial_r\log c_2^2\,H_{tt}'\,-\,\partial_r\log (c_1 c_2) (H_{xx}'+H_{yy}')\,=\,
{c_3^2\over c_1^2}\,\Big(\omega^2(H_{xx}+H_{yy})\,+\,2\omega q\,H_{ty}\Big)\,+\,\rc\rc
&&\qquad\qquad\qquad\qquad\qquad
+q^2\,{c_3^2\over c_2^2}\,(H_{tt}-H_{xx})
+c_3^2\,\hat\delta V\,-\,\sum_{\Psi}\,{\Psi'\over \alpha_{\Psi}}\,\hat \Psi\,\,.
\label{first_order_metric_fluct}
\eear
In (\ref{second_order_metric_fluct}) $W$ is the function defined in (\ref{W_def}), which vanishes in our background and, therefore, will be omitted from now on. The function $\tilde W$ appearing in the first equation in (\ref{second_order_metric_fluct}) is defined as:
\beq
\tilde W\,=c_3^2\,\,V\,+\,2\,\partial_r^2\log c_1\,+\,
 2\,\partial_r\log c_1\,\partial_r
\log\Big({c_1\,c_2^2\over c_3}\Big)\,\,.
\label{tilde_W_def}
\eeq
This function also vanishes in our background and will also  be omitted in the equations that follow. 

We now write the equations for the scalar fluctuations in terms of the new fields $Z_{\hat\Psi}$ defined 
in (\ref{Z_Psi_def}). With this aim, let us define ${\cal W}_{\phi}$, ${\cal W}_{\gamma}$, ${\cal W}_{\lambda}$ and ${\cal W}_{\beta}$ as
 the following linear combinations of the $Z_{\hat\Psi}$'s:
\beq
{\cal W}_{\Psi}\,=\,\alpha_{\Psi}\,\sum_{\Psi'}\,
{\partial^2 V\over \partial \Psi\,\partial\Psi'}\,\,
Z_{\hat\Psi'}\,\,.
\eeq
It turns out that the  equations of motion of the scalar fluctuations can be written as:
\bear
&&Z_{\Phi}''\,+\,\partial_r\,\log\Big({c_1 c_2^2\over c_3}\Big)\,Z_{\Phi}'\,+\,
c_3^2\Big({\omega^2\over c_1^2}\,-\,{q^2\over c_2^2}\Big)\,Z_{\Phi}\,-\,
{3\,c_3^2\over 7}\big({\cal W}_{\phi}-2{\cal W}_{\beta}\big)\,=\,0\,\,,\rc\rc
&&Z_{\Gamma}''\,+\,\partial_r\,\log\Big({c_1 c_2^2\over c_3}\Big)\,Z_{\Gamma}'\,+\,
c_3^2\Big({\omega^2\over c_1^2}\,-\,{q^2\over c_2^2}\Big)\,Z_{\Gamma}\,-\,
\,c_3^2\,{\cal W}_{\gamma}\,=\,0\,\,,\rc\rc
&&Z_{\Lambda}''\,+\,\partial_r\,\log\Big({c_1 c_2^2\over c_3}\Big)\,Z_{\Lambda}'\,+\,
c_3^2\Big({\omega^2\over c_1^2}\,-\,{q^2\over c_2^2}\Big)\,Z_{\Lambda}\,-\,
\,c_3^2\,{\cal W}_{\lambda}\,=\,0\,\,,\rc\rc
&&Z_{B}''\,+\,\partial_r\,\log\Big({c_1 c_2^2\over c_3}\Big)\,Z_{B}'\,+\,
c_3^2\Big({\omega^2\over c_1^2}\,-\,{q^2\over c_2^2}\Big)\,Z_{B}\,-\,
{2\,c_3^2\over 7}\big(2{\cal W}_{\beta}-{\cal W}_{\phi}\big)\,=\,0\,\,.\qquad\qquad
\label{eoms_Zs}
\eear
By combining the first and last equations in (\ref{eoms_Zs}),  one can immediately show that  the scalar $Z_S$ 
defined in (\ref{Z_S_def}) satisfies the simple equation:
\beq
Z_{S}''\,+\,\partial_r\,\log\Big({c_1 c_2^2\over c_3}\Big)\,Z_{S}'\,+\,
c_3^2\Big({\omega^2\over c_1^2}\,-\,{q^2\over c_2^2}\Big)\,Z_{S}\,=\,0\,\,.
\label{ZS_eom}
\eeq
More explicitly, this equation can be written as:
\beq
Z_{S}''\,+\,\partial_r\,\log\big(r^{{13\over 3}}\,b(r)\big)\,Z_{S}'\,+\,
{R^4\over r^4\,b^2(r)}\,\big(\omega^2\,-\,b(r)\,q^2\big)\,Z_{S}\,=\,0\,\,.
\label{ZS_eom_explicit}
\eeq

One can demonstrate that the equation  for  the gauge invariant metric fluctuation $Z_H$ takes the form:
\beq
Z_{H}''\,+\,{\cal F}(r)\,Z_{H}'\,+\,{\cal G}(r)\,Z_H\,+{\cal H}(r)\,Z_S\,=\,0\,\,,
\label{ZH_eom}
\eeq
where the functions ${\cal F}(r)$, ${\cal G}(r)$ and ${\cal H}(r)$ are given by:
\bear
&&
{\cal F}(r)\,=\,\partial_r\,\log\Big({c_1\,c_2^2\over c_3}\Big)\,-\,4\,\partial_r\log\Big({c_1\over c_2}\Big)\,+\,
\xi_1(r)\,\,, \rc\rc
&&
{\cal G}(r)\,=\,c_3^2\Big({\omega^2\over c_1^2}\,-\,{q^2\over c_2^2}\Big)\,+\,
4\Big[\partial_r \log\Big({c_1\over c_2}\Big)\Big]^2\,-\,\partial_r \log\Big({c_1\over c_2}\Big)\,\xi_1(r) \,\,,
\rc\rc
&&
{\cal H}(r)\,=\,-{q^2\over \omega^2}\,{c_1^2\over c_2^2}\,\Big[
\partial_{\beta} V\,\Big(1-{\partial_r\log c_1\over \partial_r\log c_2}\Big)\,c_3^2\,+\,
6\,\beta'\,\xi_2(r)\Big]\,\,,
\eear
with:
\bear
&&\xi_1(r)\,=\,{ q^2 \partial_r c_1^2\,\,{\partial^2_r\log c_2\over (\partial_r\log c_2)^2}
\Big(1-{\partial^2_r\log c_1\, \partial_r\log c_2\over \partial^2_r\log c_2 \,\partial_r\log c_1}\Big)
+4\, \omega^2\,\partial_r c_2^2\,\,\Big(1-{\partial_r \log c_1\over \partial_r\log c_2}\Big)
\over
q^2\,c_1^2\,\Big({\partial_r\log c_1\over \partial_r\log c_2}+1\Big)\,-\,2\omega^2\,c_2^2}\,\,,
\rc\rc\rc
&&\xi_2(r)\,=\,{
(q^2 c_1^2-\omega^2 c_2^2){
\partial_r^2\log c_2\partial_r\log c_1-\partial_r^2\log c_1\partial_r\log c_2\over
\partial_r\log c_2}+2\omega^2 c_2^2(\partial_r\log c_1-\partial_r\log c_2)^2
\over q^2\,c_1^2\partial_r\log (c_1 c_2)\,-\,2\omega^2 c_2^2\,\partial_r\log c_2}\,\,.
\qquad\qquad
\eear
Notice that $Z_H$ only couples to the scalar field $Z_S$.  Thus, we are left with (\ref{ZS_eom}) and (\ref{ZH_eom}) to be solved in the hydrodynamic approximation. 

The scalar fluctuation equation (\ref{ZS_eom_explicit}) only involves the function $Z_S$ and, therefore, can be studied independently of $Z_H$. Let us  adopt the following ansatz for $Z_S(r)$ :
\beq
Z_S(r)\,=\,\big[ b(r)\big]^{-{i\hat\omega\over 2}}\,\,K(r)\,\,,
\eeq
for which the infalling boundary conditions at the horizon are satisfied if $K(r)$ is regular at the horizon.  Actually, it is much more convenient to change variables and work in the  variable $x$ defined in (\ref{x_r_def}). 
Recall that  the horizon is located at $x=0$, whereas the boundary is at $x=1$. The fluctuation equation (\ref{ZS_eom_explicit}) is equivalent to the following equation for $K(x)$:
\beq
K''\,+\,{1-2i\,\hat\omega\over x}\,\,K'\,+\,
{\big[1\,-\,(1-x^2)^{{7\over 5}}\,\big]\hat\omega^2\,-\,x^2\,\hat q^2\over
x^2\,(1-x^2)^{{7\over 5}}}\,K\,=\,0\,\,,
\label{eom_K(x)}
\eeq
where now the primes  denote derivatives with respect to the new variable $x$. We want to solve (\ref{eom_K(x)}) for low $\hat q$. Accordingly, we expand $K(x)$ as:
\beq
K(x)\,=\,K_0(x)\,+\,i\hat q\,K_1(x)\,+\,\hat q^2\,K_2(x)\,\,.
\label{K_q_expansion}
\eeq
Plugging (\ref{K_q_expansion}) and (\ref{dispersion})  into (\ref{eom_K(x)}) and separating the different orders in $\hat q$, we find the following systems of equations:
\bear
&&K_0''+{K_0'\over x}\,=\,0\,\,,\rc\rc
&&K_1''+{K_1'\over x}\,=\,{2 v_s\over x}\,K_0'\,\,,\rc\rc
&&K_2''\,+\,{K_2'\over x}\,=\,{2\,\Gamma\over x}\,K_0'\,+\,
{1\over (1-x^2)^{{7\over 5}}}\,\Big(1-{v_s^2\over x^2}\Big)K_0\,+\,
{v_s^2\over x^2}\,K_0\,-\,{2 v_s\over x}\,K_1'\,\,.
\label{K0_K1_K2_eqs}
\eear
The equation for $K_0(x)$ can be straightforwardly integrated in general:
\beq
K_0(x)\,=\,c_1+c_2\,\log x\,\,,
\eeq
where $c_1$ and $c_2$ are constants. The regularity requirement of $K_0(x)$ at $x=0$ implies that $c_2=0$, while the condition  $K_0(x=1)=0$ imposes that $c_1$ vanishes and, thus $K_0(x)=0$. For this value of $K_0(x)$ the equation
for $K_1(x)$ in (\ref{K0_K1_K2_eqs}) is the same as the one for  $K_0(x)$. Therefore, the only valid solution for our boundary conditions is $K_1(x)=0$. Furthermore, the same happens for 
$K_2(x)$ and, thus, we finally have that the solution for $K(x)$ satisfying the boundary conditions is the trivial one, namely:
\beq
K(x)\,=\,0\,\,.
\eeq
Therefore, it follows that $Z_S(r)=0$, as claimed in the main text.

\end{document}